\documentclass[12pt,preprint]{aastex}
\usepackage{color}

\shorttitle{Stereoscopic polar plume reconstructions}
\shortauthors{Feng et al.}

\begin{document}
\title{Stereoscopic polar plume reconstructions from
            STEREO/SECCHI images}

\author{L. {Feng}\altaffilmark{1,2},
        B. {Inhester}\altaffilmark{1},
        S.K. {Solanki}\altaffilmark{1},
        K. {Wilhelm}\altaffilmark{1},
        T. {Wiegelmann}\altaffilmark{1},
        B. {Podlipnik}\altaffilmark{1},
        R.A. {Howard}\altaffilmark{3},
        S.P. {Plunkett}\altaffilmark{3},
        J.P. {Wuelser}\altaffilmark{4},and
        W.Q. {Gan}\altaffilmark{2}
        }

\altaffiltext{1}{Max-Planck-Institut f\"ur Sonnensystemforschung,
       Max-Planck-Str.2, 37191 Katlenburg-Lindau, Germany}
\altaffiltext{2}{Purple Mountain Observatory,
  Chinese Academy of Sciences, 210008 Nanjing, China}
\altaffiltext{3}{Naval Research Laboratory, Code 7660, 4555 Overlook Ave. SW,
  Washington D.C., USA, 20375}
\altaffiltext{4}{Solar and Astrophysics Lab., Lockheed Martin ATC,
  3251 Hanover St., Palo Alto, CA 94304, USA}

\begin{abstract}

We present stereoscopic reconstructions of the location and inclination of
polar plumes of two data sets based on the two simultaneously recorded images
taken by the EUVI telescopes in the SECCHI instrument package onboard the
\emph{STEREO (Solar TErrestrial RElations Observatory)} spacecraft.
The ten plumes investigated show a superradial expansion in the coronal hole
in 3D which is consistent with the 2D results. Their deviations from the
local meridian planes are rather small with an average of $6.47^{\circ}$.
By comparing the reconstructed plumes with a dipole field with its axis
along the solar rotation axis, it is found that plumes are inclined more
horizontally than the dipole field. The lower the latitude is, the larger is
the deviation from the dipole field. The relationship between plumes and
bright points has been investigated and they are not always associated. For
the first data set, based on the 3D height of plumes and the electron density
derived from SUMER/\emph{SOHO} Si\,{\sc viii} line pair, we found that
electron densities along the plumes decrease with height above the solar surface.
The temperature obtained from the density scale height is 1.6 to 1.8
times larger than the temperature obtained from Mg\,{\sc ix} line ratios. We
attribute this discrepancy to a deviation of the electron and the ion
temperatures. Finally, we have found that the outflow speeds
studied in the O\,{\sc vi} line in the plumes
corrected by the angle between the line of sight and the plume orientation
are quite small with a maximum of 10 $\mathrm{km~s^{-1}}$. It
is unlikely that plumes are a dominant contributor to the fast solar wind.

\end{abstract}

\keywords{solar corona, plume, stereoscopy}

\section{Introduction}

With the launch of NASA's \emph{STEREO} mission in October 2006, a new
dimension has been added to solar coronal observations. The two spacecraft
orbit the Sun and separate slowly in longitude by about $45^{\circ}$ per year.
The first year after launch is the best time for the 3D reconstruction
of coronal structures, such as magnetic loops
\citep{Wiegelmann:Inhester:2006,Feng:etal:2007a,Feng:etal:2007b,Aschwanden:etal:2008a},
polar plumes \citep{Curdt:etal:2008}, polar jets \citep{Patsourakos:etal:2008}, solar
prominences (filaments), etc., which can be achieved by analysing the image pairs
taken by the EUVI telescopes that are part of the SECCHI suite of imagers.
EUVI is very similar to EIT/\emph{SOHO}, but with
higher spatial resolution ($1.6^{\prime\prime}$ per pixel) and larger field of view
(out to 1.7 $R_\odot$). For the objectives of the mission and more details about
the EUVI telescopes see \citet{Howard:etal:2008} and \citet{Wuelser:etal:2004}.

As the solar activity was low in 2007, the appearance of the solar corona
over the poles was dominated by extended raylike features, the so-called polar plumes.
They are believed to trace out open magnetic field lines and have been intensively
analysed in white light, EUV and soft X-rays. Coordinated observations using
the \emph{SOHO} spacecraft and the ground-based HAO (High Altitude Observatory)
Mk-3 coronagraph
of the polar plumes from the photosphere to approximately 15 $R_\odot$
were presented by \citet{DeForest:etal:1997} and out to 30 $R_\odot$ or
45 $R_\odot$ by \citet{DeForest:etal:2001}. \citet{Gabriel:etal:2003,Gabriel:etal:2005}
have suggested that plumes are a substantial contributors to the fast
solar wind, whereas \citet{Habbal:1992}; \citet{Wang:1994} and
\citet{Wilhelm:etal:2000} have a different point of view. They suggest that
the interplume regions are the dominant source of the high-speed solar wind.
Another debate about plumes is whether they are rooted in an unipolar
magnetic field configuration \citet[e.g.][]{Newkirk:Harvey:1968,DeForest:etal:1997}
or are formed by the unipolar magnetic field reconnected with emerging bipolar regions
\citet[e.g.][]{Wang:Sheeley:1995}.

One handicap of all previous studies was the fact that plumes, even though
well visible in 2D images beyond the limb, could not be reliably traced to
the solar surface. We have reconstructed the 3D geometry of the polar plumes
by analysing the simultaneously observed EUVI images for two data sets,
April~7 and June 1, 2007. Preliminary results of the 3D plume coordinates
obtained on one of the studied days (April 7) are shown in \citet{Curdt:etal:2008}.
In this paper, we describe an improved method of stereoscopic reconstruction
in detail since in \citet{Curdt:etal:2008} the emphasis was put on the
multi-spacecraft observations from \emph{Hinode}, \emph{STEREO} and \emph{SOHO}.
Furthermore, based on the 3D coordinates, we have calculated the
inclination of plumes to the LOS of the Earth and to their local radial
direction for both date sets. We have compared the 3D orientation of plumes with
the local direction of a dipole magnetic field. The dipole field is the lowest order
approximation of the coronal magnetic field at times of low solar activity
when the plume observations were made. Additionally, we have calculated the
footpoint positions of the reconstructed plumes and projected them onto the
EUVI images to investigate their relationship to EUV bright points. For the
first data set, \emph{SOHO}/SUMER observations were also available. We have
used them to determine physical parameters, including the temperature, density
and line-of-sight (LOS) Doppler shift. By projecting a 3D plume onto the SUMER density
map, its density scale height could be calculated. The temperature
corresponding to this scale height and the temperature derived spectroscopically
from SUMER has been compared.

\section{The data}

\emph{STEREO} was launched somewhat after the solar activity minimum which provides us with
good opportunities to observe polar plumes. We selected two data sets for this
study, one in the south polar cap on 2007-04-07 22:01:17 UTC and the
other in the north polar cap on 2007-06-01 00:09:00 UTC. Both were recorded
by the two almost identical SECCHI/EUVI telescopes at $\lambda = 17.1~nm$
corresponding to a formation temperature of roughly 1 MK.
The position information of both spacecraft, the exposure time and the
compression mode are given in Table \ref{tab:orbit_plume}.

On April 7, the separation of the two spacecraft was $3.6^ \circ$.
The HEEQ (Heliocentric Earth EQuatorial coordinate system;
\citet{Thompson:2006}) latitudes of the two spacecraft were well below the
Sun's equatorial plane. At that time the solar south pole was
tilted towards the spacecraft which was very appropriate for observing as much
as possible of the southern polar area.
In addition, to improve the signal to noise ratio of the EUVI images,
a longer exposure time and a smaller compression were applied. The image
pairs chosen in this work have 20 s exposure time, compared to the normal
exposure time of 2 s. The images were compressed by the format
ICER (a wavelet-based image compression file format)~4 which requires two
times the storage of images obtained after applying the usual ICER 6
compression used ordinarily for EUVI images at 17.1~nm. In Figure \ref{fig:euvi_0407},
the southern polar cap in both EUVI views is presented with five plume pairs
marked by numbers below them.

In parallel, SUMER performed a raster scan from 2007-04-07 01:01 UTC to
2007-04-08 12:19 UTC in the southern corona. The scan direction
was from solar west to east. The details of the SUMER observations can be
found in \citet{Curdt:etal:2008}. To combine with 3D plume
geometry, we derived the electron density map from the emission line
pair Si\,{\sc viii} at 144.0 nm and 144.6 nm, the electron temperature map
from line pair Mg\,{\sc ix} at 70.6 nm and 75.0~nm, and the LOS
Doppler shift map from the O\,{\sc vi} at 103.2 nm and 103.8 nm.

The second analysed EUVI data set was recorded on June 1, when the separation
angle between the spacecraft was $10.6^\circ$. This increased separation angle
reduced the reconstruction uncertainty considerably. At the same time, this
separation angle is still sufficiently small so that the correspondence between
individual plumes in both EUVI images can still be unambiguously identified.
The related spacecraft positions, observation and compression
parameters are listed on Table \ref{tab:orbit_plume}.

\section{The reconstruction}

\begin{figure*} 
  \includegraphics[width=15cm,height=15cm]{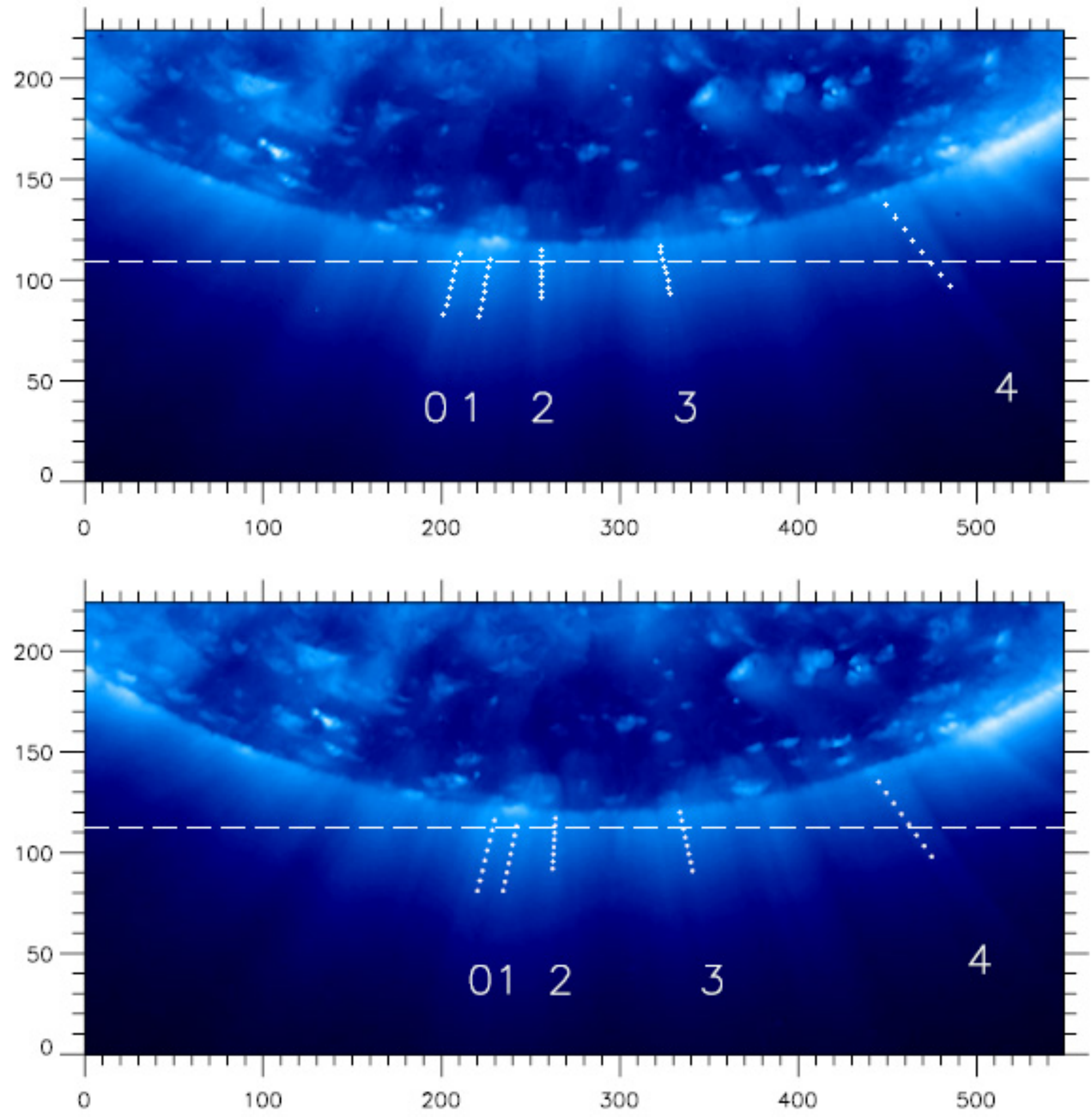}
  \caption{The south polar cap observed on
  2007-04-07 22:01:17 UTC at $\lambda$ = 17.1 nm
  by EUVI A (upper) and B (bottom). The corresponding epipolar lines are
  approximated by the two long-dashed lines. The dotted lines
  are the identified plumes (see Section 3 for details).}
  \label{fig:euvi_0407}
\end{figure*}

To reconstruct the 3D geometry of the plume, the first step is to identify
the points along the plume axis and associate
the corresponding plume pair in the two simultaneous EUVI images.
To establish this correlation between the plumes, we extracted from each image
the radiance profiles along corresponding epipolar lines \citep{Inhester:2006}
in both images. Figure~\ref{fig:inten_epi_plot} gives an example for the April 7 data.
The radiance distributions along the corresponding epipolar lines, which are
approximated by the two long-dashed lines in Figure \ref{fig:euvi_0407}, were
smoothed by taking the running mean of the radiance over three pixels. The plume centers
were selected according to the local radiance maxima and marked by the
corresponding numbers.

For plumes 0, 1, 2 and 4 the association is clear. However for plume 3 there
is some uncertainty due to the complicated superposition of probably several
plumes along the line of sight. The radiance distribution around plume 3 in EUVI A is
dominated by one prominent peak (marked as 3a in Figure \ref{fig:inten_epi_plot})
with a smaller peak (3) on the right side, whereas in EUVI B two distinct
peaks (3b and 3) appear. There are four possible ways to associate the two
plume signatures in the two images. Trying all combinations, we judged their
likelihood from the inclination of the resulting 3D plumes. When a point on
the 3D plume leaves from its footpoint on the solar surface, its distance to
the solar rotation axis should increase as well. The second criterion is that
the angle between the plume and the local meridian
plane passing through the plume footpoint should be as small as possible
(see Figure \ref{fig:plume_geometry}). In this way, we find that the
combination of peak 3 in image A and peak 3 in image B gave the most
reasonable result. In Figure \ref{fig:cross_sec}, we sketch the
situation which we think yields the different peaks in the two
images for plume 3. Here we assumed a circular shape and a uniform
radiance distribution within the cross section of two plumes of different
radii. Figure \ref{fig:cross_sec} shows the resulting radiance
distributions from two different perspectives.

\begin{figure*} 
  \hspace*{\fill}
  \includegraphics[width=15cm,height=15cm]{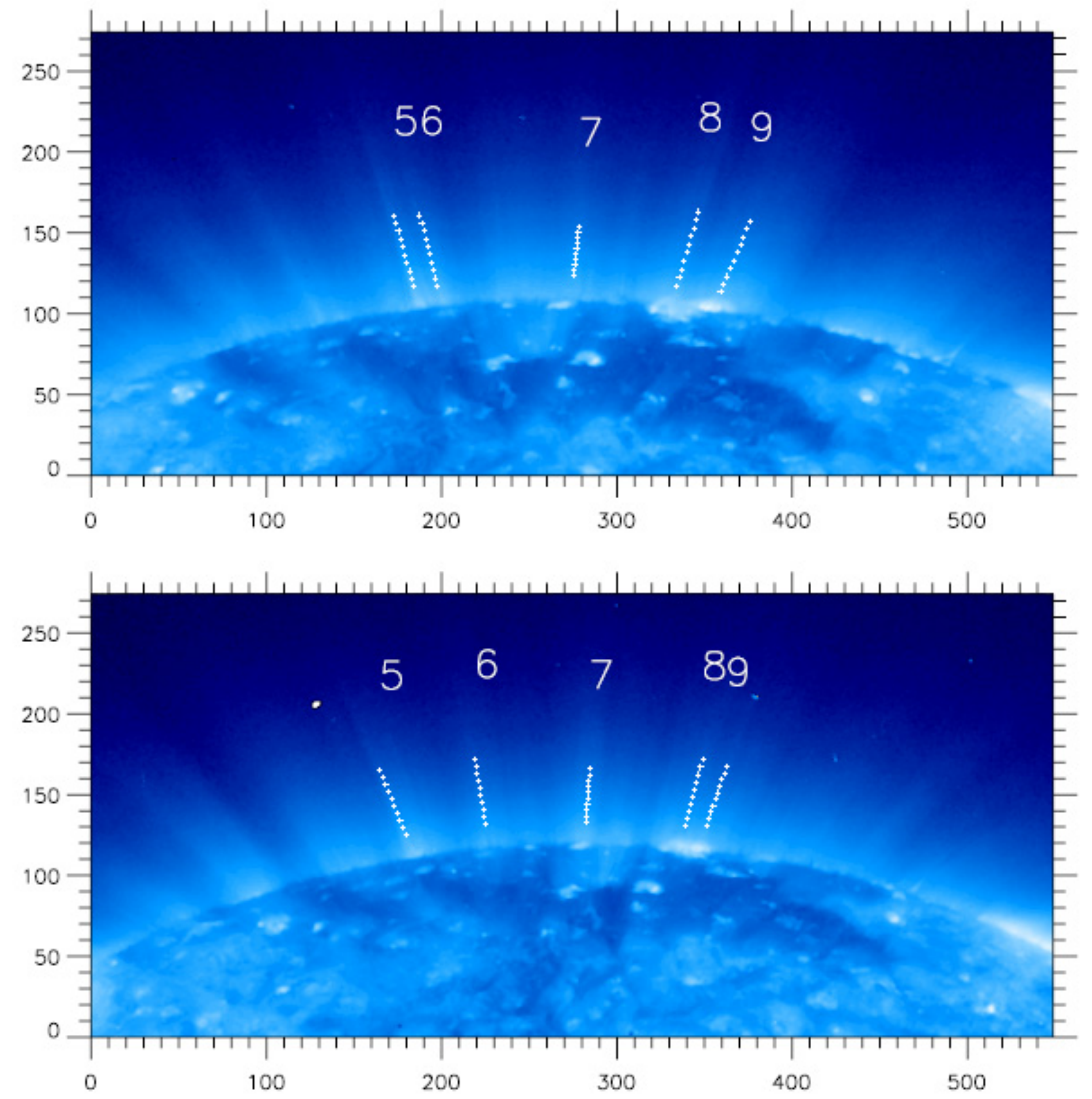}
  \hspace*{\fill}
  \caption{The north polar cap observed at 2007-06-01 00:09:00 UTC
  by EUVI A (upper) and B (bottom).}
  \label{fig:euvi_0601}
\end{figure*}

\begin{figure} 
  \includegraphics[height=6cm, width=8cm]{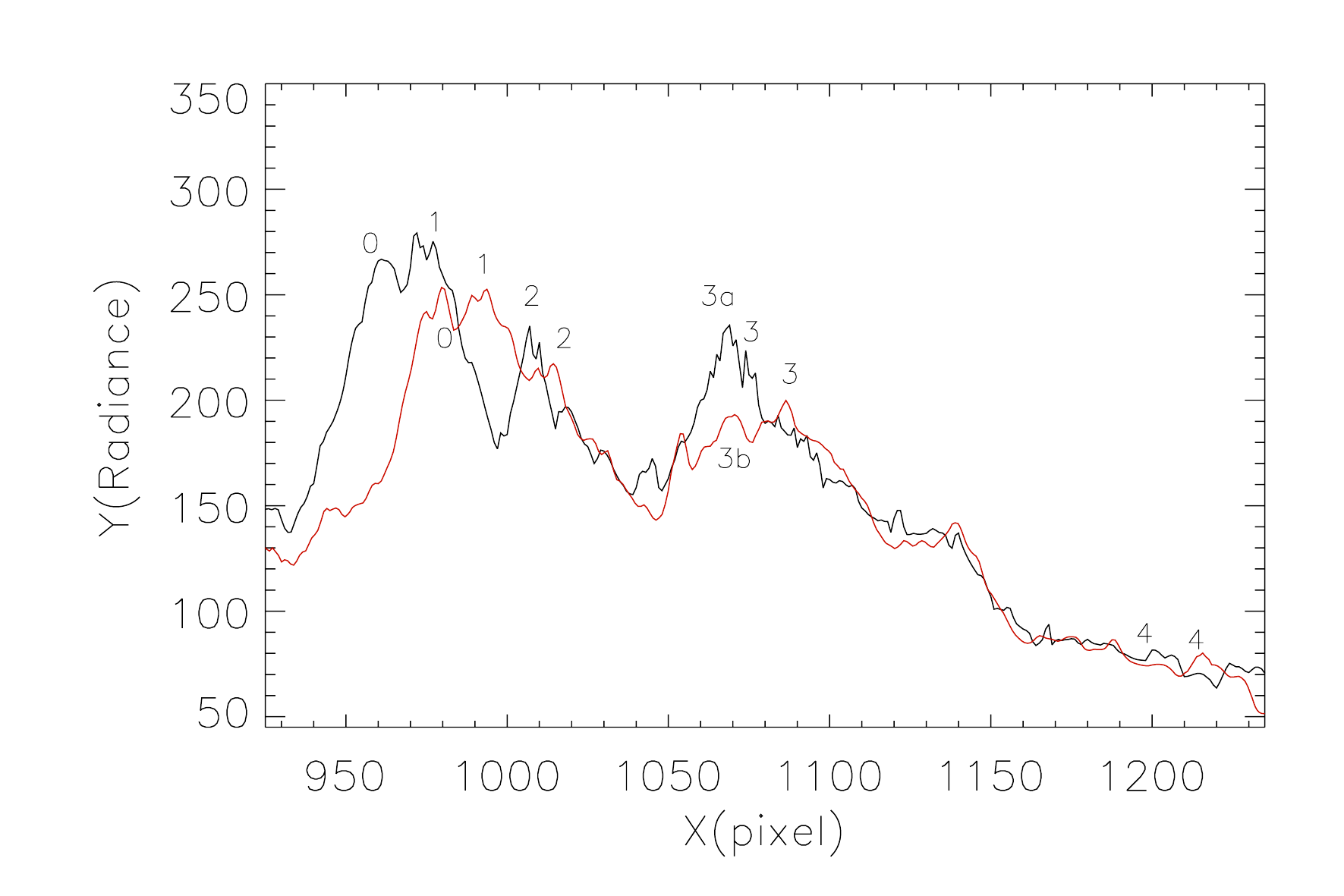}
  \caption{
  The radiance distributions smoothed over three pixels along the corresponding
  epipolar lines in each EUVI image observed in April. The black and red
  lines are for EUVI A and B,
  respectively. The $x$ coordinate of the solar center in A and B lies at 1020.62
  and 1035.52 in units of pixels, respectively.
  The numbers refer to the plumes in Figure \ref{fig:euvi_0407}.}
  \label{fig:inten_epi_plot}
\end{figure}

\begin{figure} 
  \includegraphics[height=6cm, width=8.5cm]{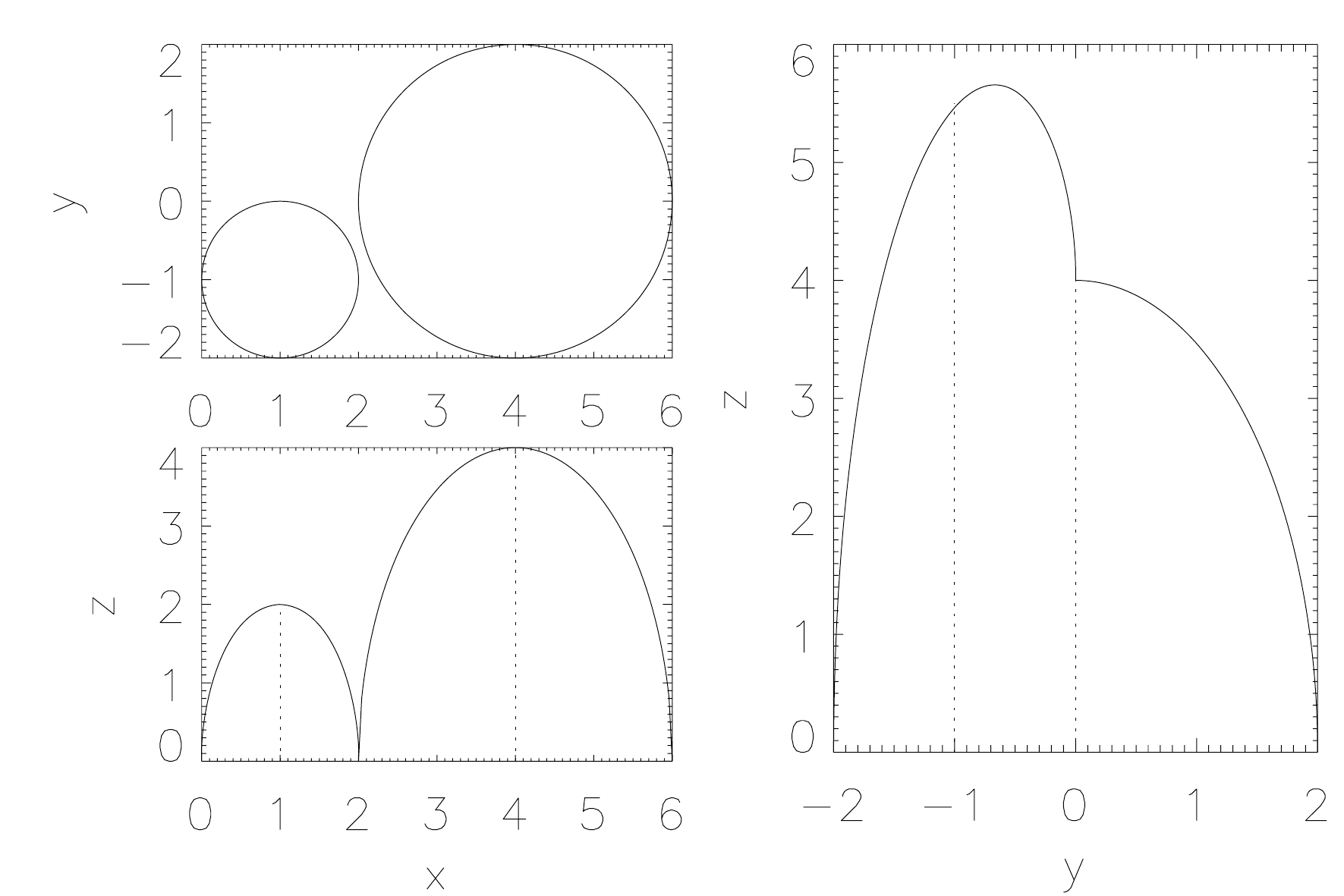}
  \caption{A schematic explanation of why a different number of peaks might be
  visible from different perspectives. The upper left panel shows the assumed cross
  section of two plumes. The lower left panel displays the LOS integrated radiance
  as seen in the $y$-direction, The right panel exhibits the radiance as seen
  along the $x$-direction.}
  \label{fig:cross_sec}
\end{figure}

Finally, the axes of the five plumes were traced by repeating the procedure
for each plume at eight different heights above the solar limb. Since in EUV
images plumes are shaped as nearly straight lines, each plume axis was approximated
by a linear function and plotted as a dotted line in Figure \ref{fig:euvi_0407}.
The identification of the plumes for the second data set observed in June 2007
is similar and the results are overplotted on the corresponding EUVI A and B
images shown in Figure~\ref{fig:euvi_0601}.

From these linear 2D plume positions in both EUVI images, we reconstruct the
3D plume locations based on epipolar geometry in the frame of the HEEQ coordinate
system. The reconstructed plumes are straight lines in 3D. In order to obtain
an error estimate, we here assume a maximal uncertainty of 3 Mm, corresponding
to an positional error in the images of $4.3^{\prime\prime}$ in EUVI A and
$4.0^{\prime\prime}$ in EUVI B, in the estimated plume position along the
epipolar line due to the three-pixel smoothing as mentioned before.
We shall see in the next section that this uncertainty is propagated to an
analogous uncertainty of the stereoscopic reconstruction.

\section{Results}

We present the different perspectives of plumes for two data sets derived
from stereoscopic reconstructions in Section~\ref{sec:stereoscopic_res}, as
well as the plume orientation analyses, plume width calculations and the
relationship between plumes and EUV bright points. In
Section~\ref{sec:stereosc_SUMER} the results obtained by combining
stereoscopic reconstructions and the plume density and temperature deduced from
SUMER observations are shown. The density scale height and its corresponding
temperature are calculated by assuming the plume in our study is in
hydrostatic equilibrium.

\subsection{Stereoscopic results}
\label{sec:stereoscopic_res}

\subsubsection{Side view and top view}

In Figure~\ref{fig:side_view}, we present a view of the 3D placement and
direction of polar plumes from a perspective that is $90^{\circ}$ to the left and
$20^{\circ}$ up compared to the view direction of \emph{STEREO A}. Of the five
reconstructed plumes in April, three are in front of the solar limb as seen
from \emph{STEREO A}, the other two are behind the limb. For the data set in
June, only plume~6 lies in front of solar limbs as seen by both spacecraft.
The black solid lines indicate the reconstruction uncertainties
\citep{Inhester:2006} calculated by assuming a maximal 3 Mm variation
of the plume axis position. The resulting uncertainties are directed mainly
half way between the view directions of \emph{STEREO A} and \emph{B}, and
are considerable for the data set in April since the spacecraft separation angle was
small at that time. With the increased separation angle in June, the
uncertainties are greatly reduced for the same assumption of 3 Mm uncertainty
in 2D.

The polar view of the ten plumes projected onto the solar equatorial plane is
shown in Figure~\ref{fig:top_view}. Larger symbols indicates the plume
positions at greater heights above the solar surface. All ten plumes oriented
close to their local meridian planes and inclined away from the rotation axis.

\subsubsection{Plume's orientation analysis}

In Figure~\ref{fig:plume_geometry} a sketch of the 3D plume geometry is
presented to analyse the plume orientation. The relevant results is shown in
Table~\ref{tab:parameter}. Concerning the latitudes and longitudes of the
plume footpoints, all the plumes are located within a latitude cone of
$20^\circ$ around the pole. For a better estimate of the outflow speeds along the plumes
from FUV (Far Ultraviolet)/ EUV spectral observations by SUMER and UVCS
on board SOHO, the angle $\beta$
between the LOS from the Earth and the plumes' orientation were calculated.
They range from $65.7^{\circ}$ to $128.6^{\circ}$ for plume 3 and 8 being almost
perpendicular to the view direction of the Earth. For the angle $\gamma$ off
the meridian plane, we found for all ten plumes a maximum departure of
$14.2^{\circ}$. This indicates that the magnetic azimuthal component $B_\varphi$ was
very small on the polar cap during the time of our observations. The deviation
of the plume projection to the solar radial
direction $\widehat{\mathbf{e_r}}$ is calculated and represented by the angle
$\psi$. This angle in general becomes larger with increasing distance of the
footpoint from the pole as shown in the upper panel of Figure~\ref{fig:dipole_tan}.
This means that the plumes do not converge to the solar center, which
is consistent with the 2D results \citep{DeForest:etal:1997,
Fisher:Guhathakurta:1995}. They found the plumes/rays appear to diverge
radially from a point between the solar center and the respective pole.

In addition, we compared the 3D plume structure which outlines the coronal
magnetic field with the assumption of a dipole with its axis along the solar
rotation axis. In this case the plume's inclination angle $i$ and the footpoint
latitude $\lambda$ should be related by $\tan i=2\tan\lambda$ (see Page 50 of
\citet{Fowler:1990} ). From the related two
rows in Table \ref{tab:parameter} we find that this relation is not well
satisfied for each plume and $|\tan i| < 2|\tan \lambda|$ in all cases.
Therefore the magnetic field is not well approximated
by this dipole field and is more horizontal, as also shown in the middle panel
of Figure~\ref{fig:dipole_tan}. In the bottom panel we check how much plumes
deviate from the dipole field and how this deviation changes with the
latitude. We find that at lower latitudes, the plumes are more horizontal
than the dipole field. This was already noted by \citet{Saito:1965} who
used a bar magnet of finite length to fit the plumes/rays at different
distances from the Sun observed during a solar eclipse.
\citet{Banaszkiewicz:etal:1998} described a simple analytic model for the
magnetic field at solar activity minimum. A dipole and a quadruple field
were added to construct the coronal magnetic field in the solar
minimum. The reconstructed 3D plumes could be used as a reference in the polar
region to test other, more sophisticated magnetic field models
\citep{Neukirch:1995,Ruan:etal:2008}.

\subsubsection{Plume's width analysis}

Besides the orientation, we calculated the width of the isolated and
prominent plume 4 from viewpoints of EUVI A and B by fitting the radiance
profiles around this plume with Gaussian distributions. For the plume width $w$
at a given height $h$ we use the definition of \citet{Aschwanden:etal:2008b}
\begin{equation}
w(h)=\frac{\int_{x_{b1}}^{x_{b2}}[I(h,x)-b(h,x)]\mathrm{d}x}{\mathrm{max}[I(h,x)-b(h,x)]},
\end{equation}
where $x_{b1}$ and $x_{b2}$ are the $x$ coordinates of the two plume boundary
points determined by the two local minima on either side of the plume
(Figure~\ref{fig:width_method}). Here, $b(h,x)$ is the linearly varying
background radiance between $x_{b1}$ and $x_{b2}$. $I(h,x)-b(h,x)$ then denotes the background
subtracted plume radiance distribution as a function of height $h$. The
intensities associated with this plume along
two epipolar lines are shown in Figure \ref{fig:width_method} as an example.
By using the epipolar geometry, we identify the
corresponding plume point in two images and calculate the plume width from
EUVI A and EUVI B, respectively. Subsequently the plume widths along the
corresponding epipolar lines are transformed to the widths perpendicular to
the plume direction within the frame of the epipolar geometry.

For this plume, we found from both viewpoints that the width very slightly
decreases by around $10~\%$ in the height range from 20 Mm to 90 Mm in 3D.
The mean width
and standard deviation from EUVI B is ($14.0 \pm 0.9$) Mm, while from A it is
($12.7 \pm 1.2$) Mm. The two widths differ by less than 1.5\,$\sigma$ so that
this measurement is consistent with a circular plume cross section. However,
given the small separation angle, a more curtain-like structure cannot be
ruled out either. More isolated and prominent plumes need to be analysed
at large separation angles to come to a conclusion regarding the cross section of
the plumes. It should be mentioned here that we
have only considered the circular or simple noncircular cross sections for
plumes and we have not taken into account the substructure that is known to exist within
plumes \citep[e.g.,][]{DeForest:2007,DeForest:etal:1997}.
Unresolved morphology can
mimic a surprising range of other effects including modifying the inferred
density (both from photometric density estimates and from line-ratio estimates,
in different ways for the two techniques) and the observed scale height.

\subsubsection{Plume and EUV bright points}

To investigate the relationship between plumes and EUV bright points, we
projected the reconstructed 3D plumes onto two EUVI images. For the data in
April, given the small separation angle, only the projection onto EUVI A is
plotted in Figure~\ref{fig:pro_euviA_0407}. Only plume 1 is associated with
a bright point. Plumes 0 and 3 were rooted in the brighter part of the coronal
hole, but not on a bright point. Considering the evolution of polar plumes
\citep{Wang:1998, Raouafi:etal:2008}, plumes 0 and 3 were observed perhaps in the
decaying phase in which the bright points have already disappeared but the
two plumes were still visible. For the big bright point close to the limb
between plumes 3 and 4, we tried to find a plume pair but none of the possible peaks
in the two plume radiance profiles (e.g., Figure~\ref{fig:inten_epi_plot})
produced a reasonable result with a footpoint close to this bright point and
an orientation roughly along the diverging direction of the magnetic field around the pole.

\begin{figure*} 
  \vbox{
  \includegraphics[width=15cm, height=6cm]{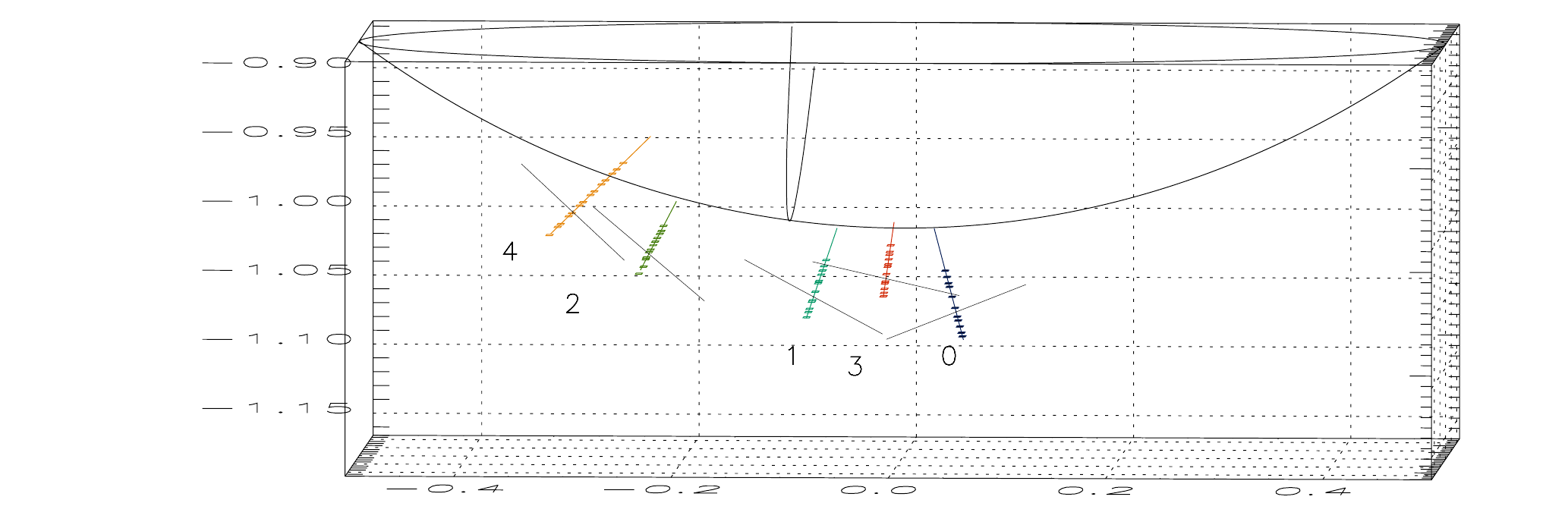}
  \includegraphics[width=15cm, height=6.5cm]{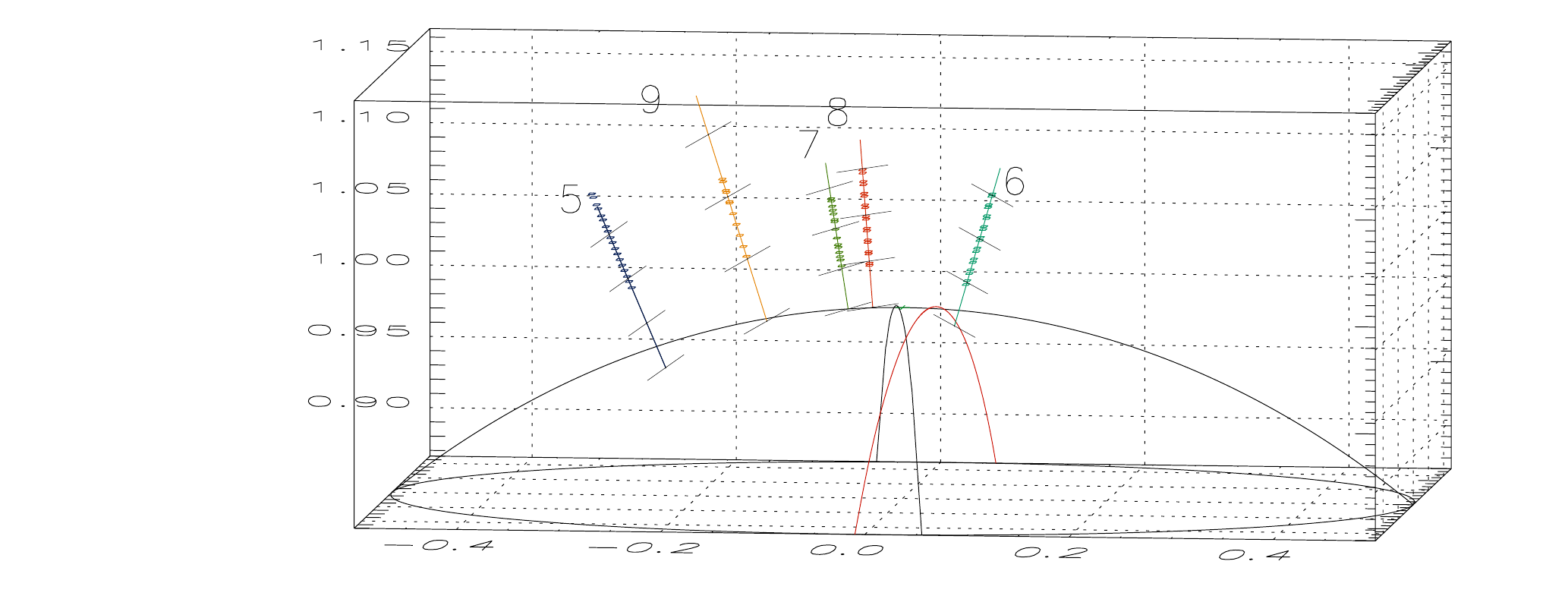}}
  \caption{Side view of the south polar cap on April 7 (upper panel) and the
  north polar cap on June 1 (lower panel): a perspective that is $90^{\circ}$
  to the left, and $20^{\circ}$ up relative to the view direction of
  \emph{STEREO} A. The coordinates $x$ and $y$ range from $-0.5~R_\odot$ to
  $0.5~R_\odot$, $z$ ranges from $-0.87~R_\odot$ to $-1.17~R_\odot$ for the
  south polar cap and from $0.87~R_\odot$ to $1.17~R_\odot$ for the north cap.
  The long curve is a circular segment crossing the pole. The shorter curves
  are the solar limbs as seen from the two spacecraft (black from \emph{STEREO A}
  and red from \emph{STEREO B}. The dotted points are the
  reconstructed 3D plume axes. The solid lines are the extrapolations back
  to $r=1~R_\odot$. The uncertainties are indicated by the black solid lines
  which are perpendicular to the plume directions in 3D.}
  \label{fig:side_view}
\end{figure*}

\begin{figure} 
  \vbox{
  \includegraphics[width=7.5cm, height=7cm]{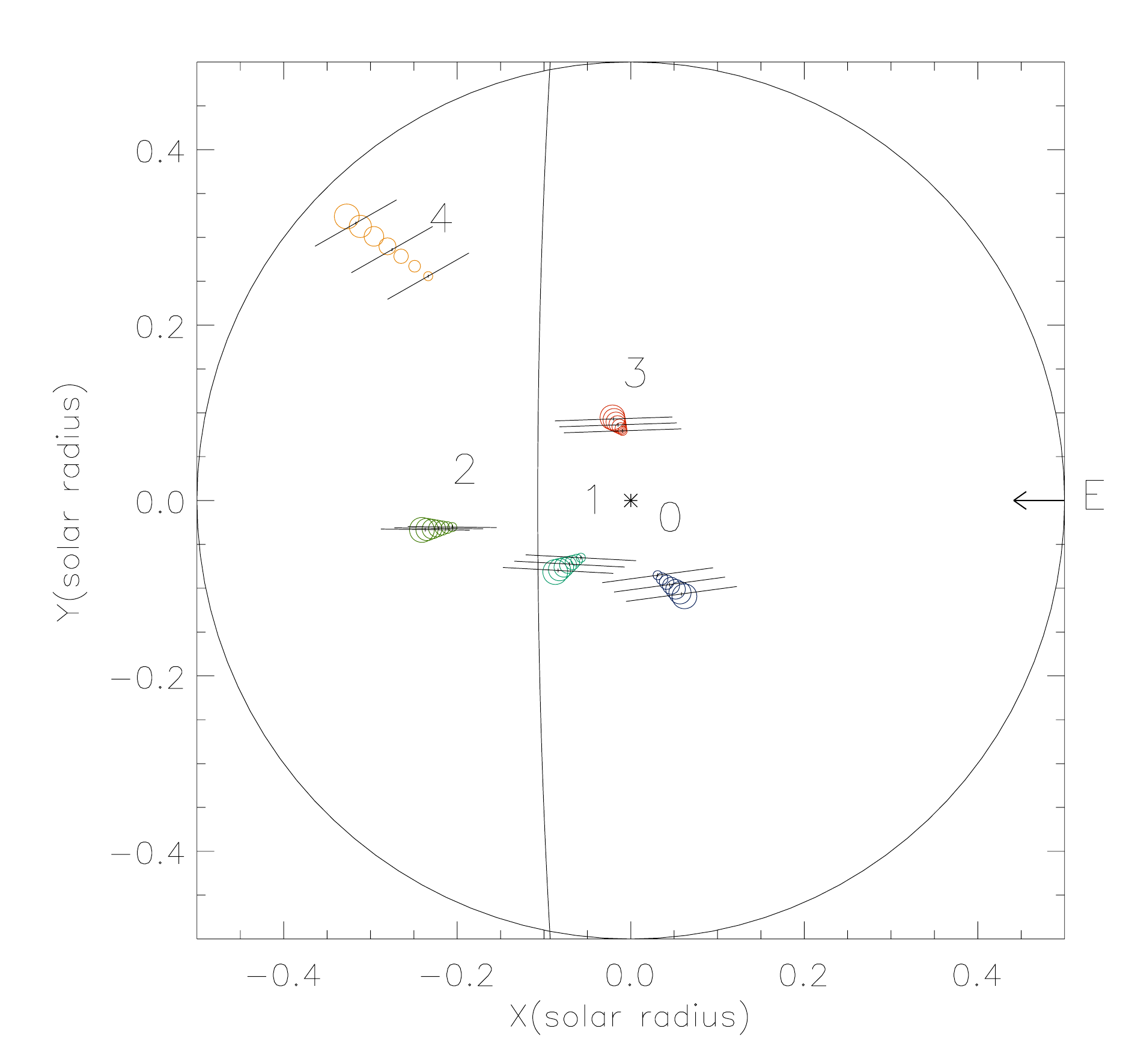}
  \includegraphics[width=7.5cm, height=7cm]{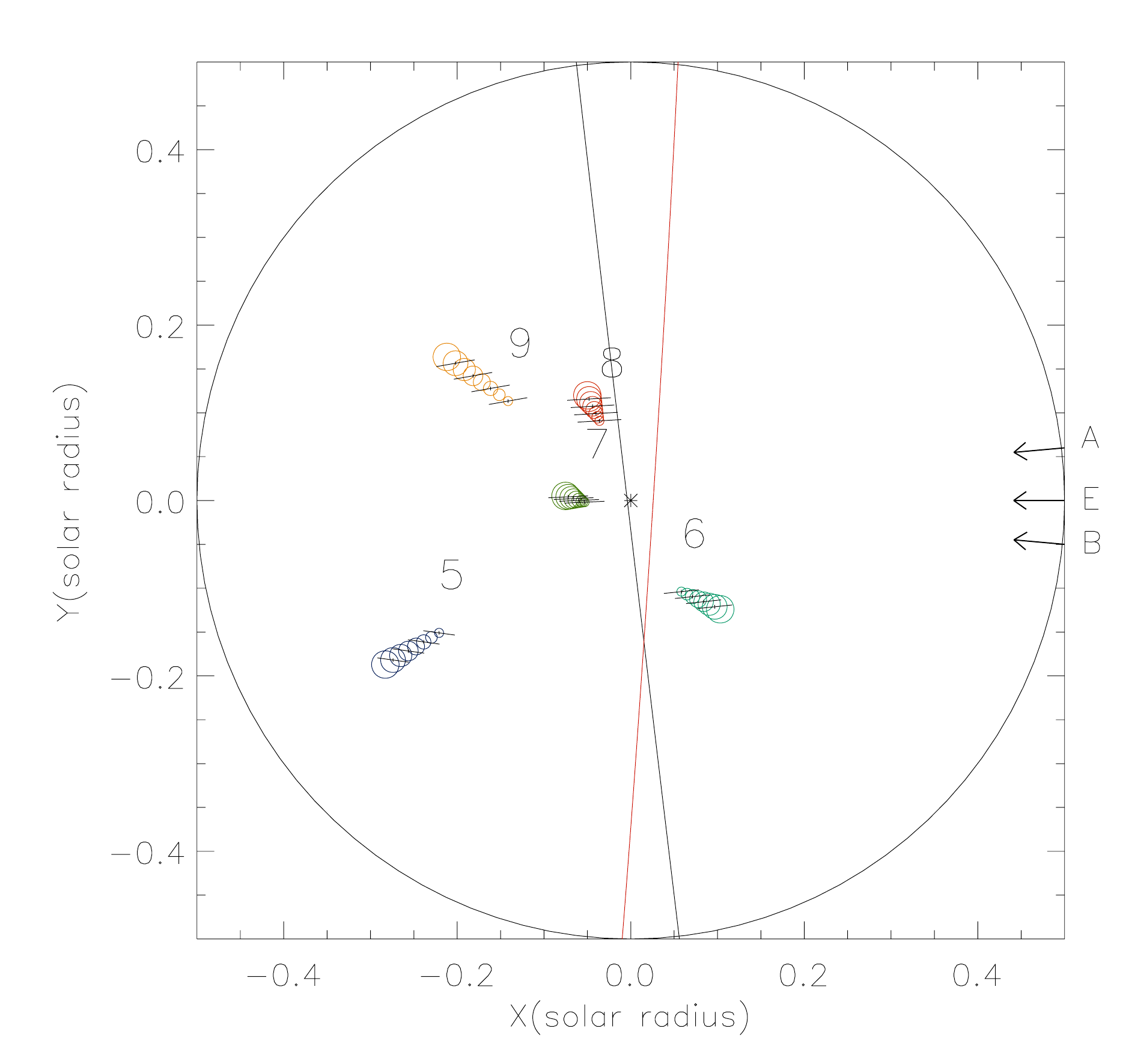}}
  \caption{Top view: projections of the reconstructed 3D plumes onto the
  solar equatorial plane in April (left) and June (right) together with the
  associated uncertainties (black solid lines). For each plume, the size of
  the circle is proportional to the distance of the 3D point to the solar
  surface. In the left panel, the solar limb as seen from the Earth is
  indicated by the curve near the pole which is marked by a star symbol. The
  view direction from the Earth indicated by the arrow marked E at the right
  edge of the figure is also shown. In the right panel, the solar limb as seen
  from \emph{STEREO A} is indicated by a black curve and seen from \emph{STEREO B}
  is indicated in red. The view directions of the two spacecraft and the
  Earth are marked on the right side.}
  \label{fig:top_view}
\end{figure}

\begin{figure} 
  \includegraphics[width=12cm,height=10cm]{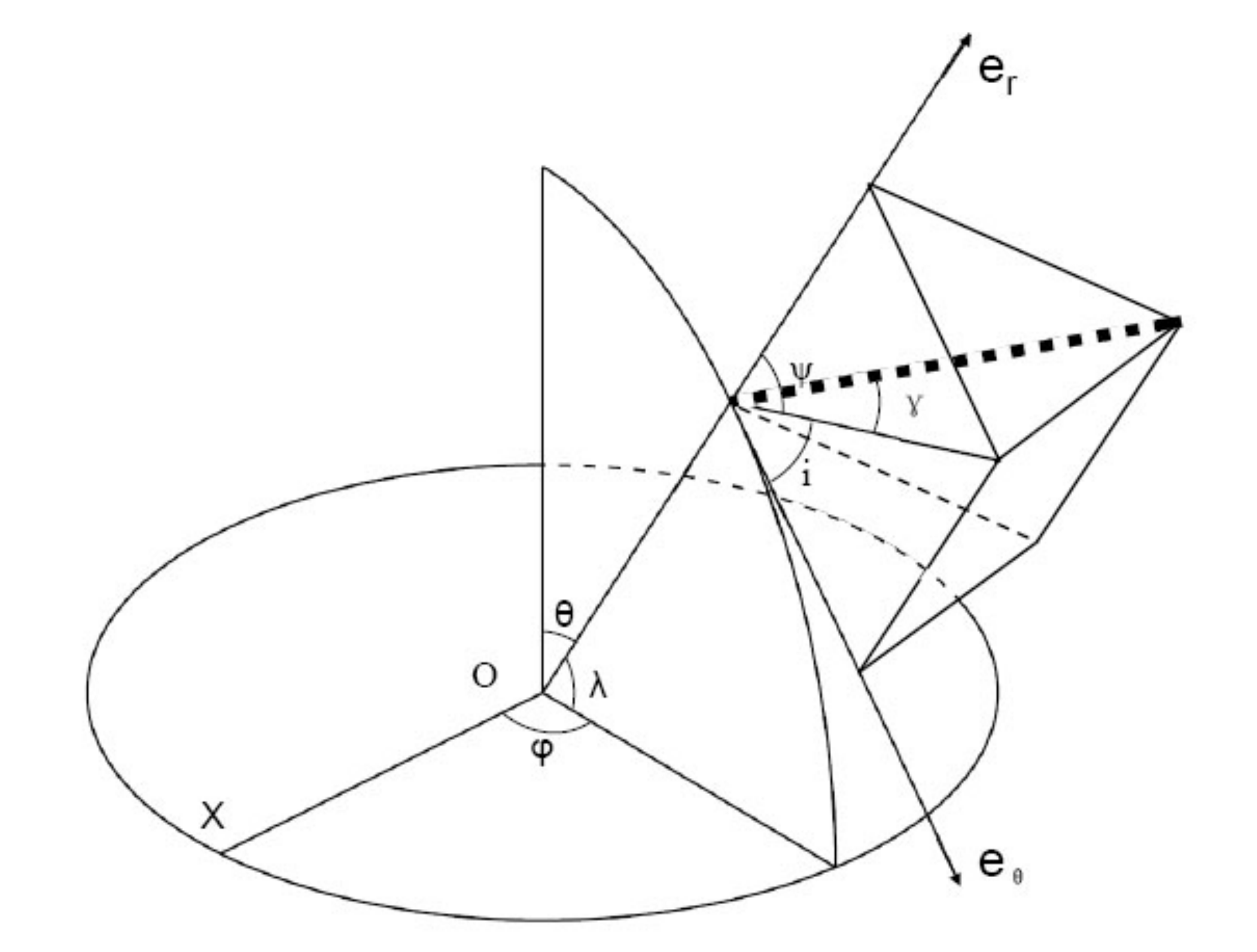}
  \caption{A schematic illustration of 3D plume geometry in the HEEQ
  coordinate system. The 3D plume is indicated by the thick dashed line
  started from the solar surface. The
  plume footpoint is parameterized with the longitude $\varphi$ and latitude
  $\lambda$ (the colatitude is denoted as $\theta$). We establish the local
  coordinate framed by $\widehat{\mathbf{e_r}}$, $\widehat{\mathbf{e_\theta}}$
  and $\widehat{\mathbf{e_\varphi}}$
  (for clearness of this figure $\widehat{\mathbf{e_\varphi}}$ is not shown)
  which originates from the footpoint.
  The projection of the 3D plume onto the local meridian plane spanned by
  $\widehat{\mathbf{e_r}}$ and $\widehat{\mathbf{e_\theta}}$ makes two angles:
  $\psi$ is the angle between the projection and $\widehat{\mathbf{e_r}}$,
  $i$ is the angle between the projection and $\widehat{\mathbf{e_\theta}}$
  and is the so-called magnetic inclination. $\gamma$ is the
  angle between the 3D plume and its projection mentioned above.}
  \label{fig:plume_geometry}
\end{figure}

\begin{figure} 
  \centering
  \includegraphics[width=8cm,height=20cm]{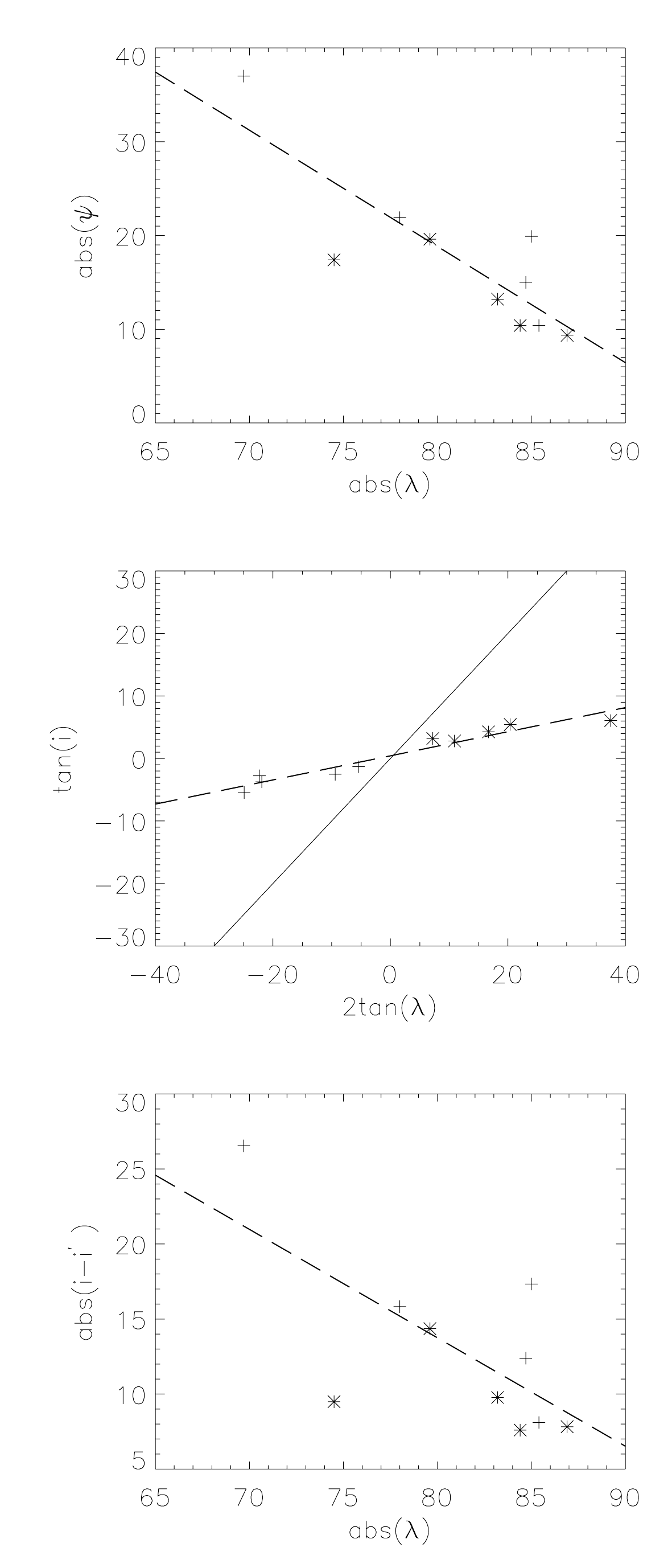}
  \caption{Upper: the absolute value of $\psi$ as a function of the absolute
  value of $\lambda$. The plus signs
  are for the data in April and star signs for the data in June.
  Middle: the plot of $\tan(i)$ versus $2\tan(\lambda)$.  The solid
  line corresponds to $\tan(i) = 2\tan(\lambda)$. Bottom: The difference of the
  magnetic inclination $i$ from the dipole field inclination $i^{\prime}$ as
  a function of the absolute latitude.}
  \label{fig:dipole_tan}
\end{figure}

\begin{figure} 
  \includegraphics[width=8.cm,height=6.0cm]{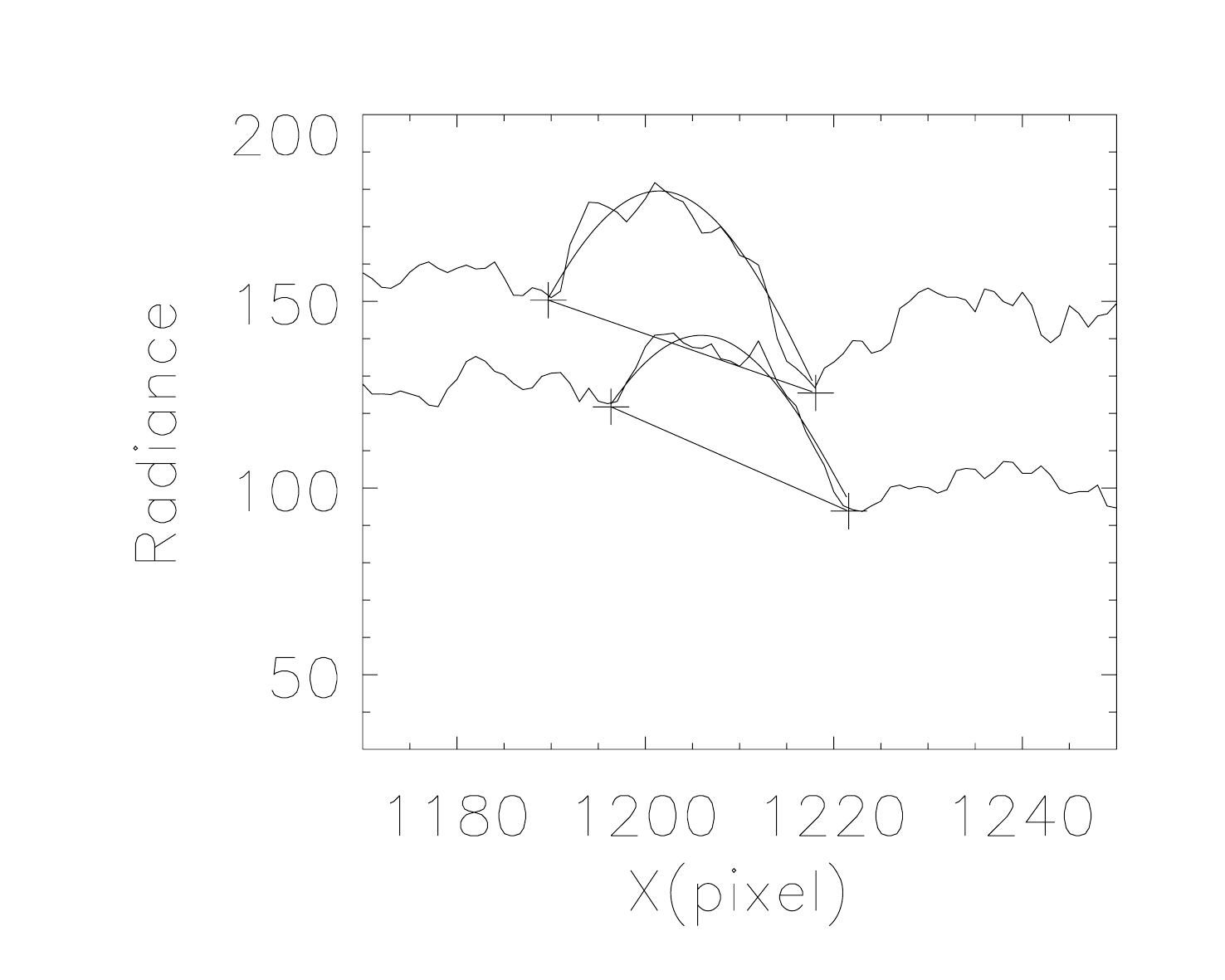}
  \caption{The radiance profile around plume 4 along two epipolar lines in
  EUVI B. On each line the two plus signs indicated the plume boundary and
  the connected
  straight line shows the plume background. Within the range between the two
  boundary points, the radiance is fit by the Gaussian distribution.}
  \label{fig:width_method}
\end{figure}

\begin{figure} 
  \includegraphics[width=8cm,height=5cm]{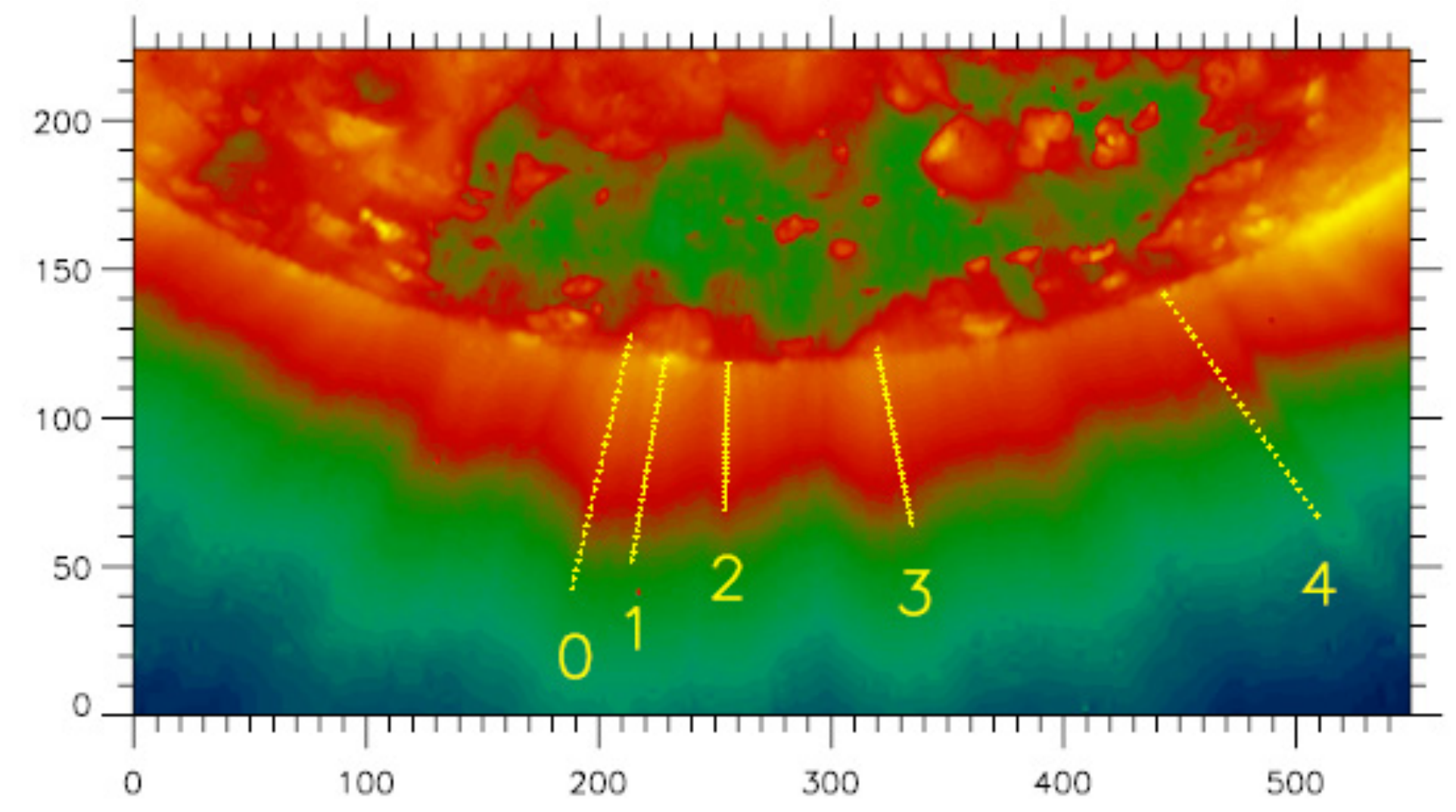}
  \caption{Projection of the 3D plumes onto the image of EUVI A observed on
          April 7.}
  \label{fig:pro_euviA_0407}
\end{figure}

\begin{figure} 
  \vbox{
  \includegraphics[width=8.cm,height=5.cm]{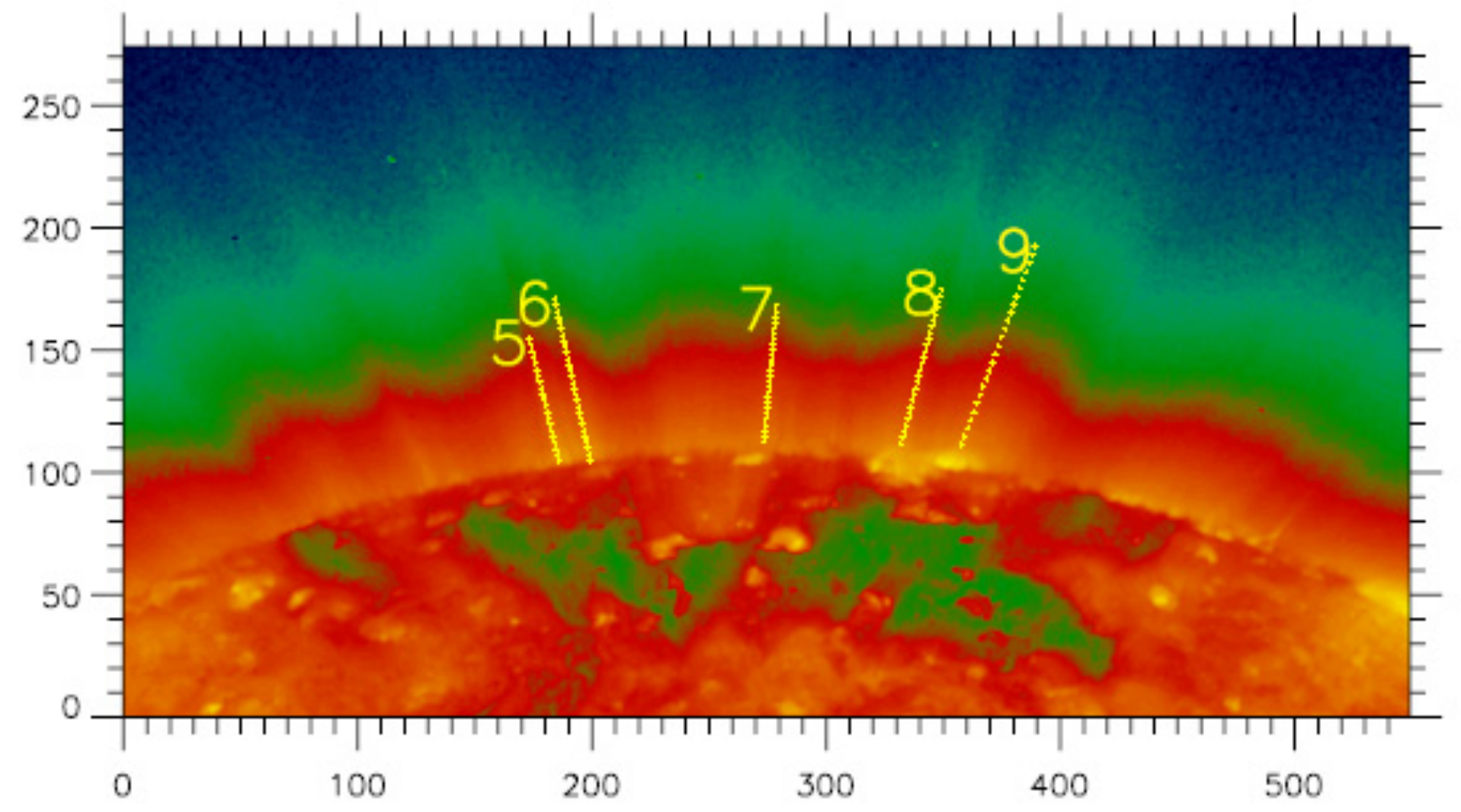}
  \includegraphics[width=8.cm,height=5.cm]{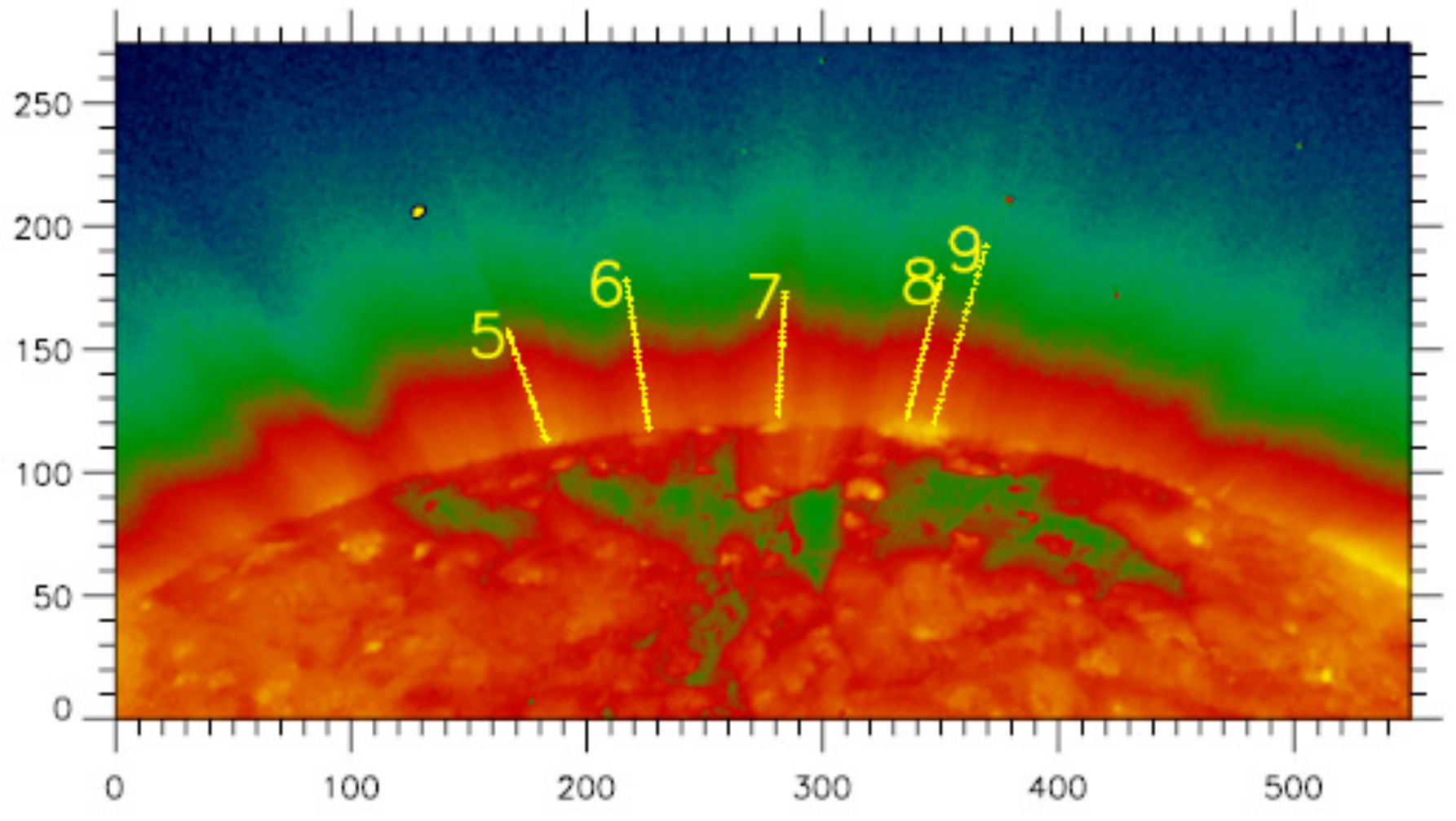}}
  \caption{Projections of the 3D plumes on EUVI A and B recorded on June 1.}
  \label{fig:proj_euvi_0601}
\end{figure}

For the data in June, Figures \ref{fig:side_view} and \ref{fig:top_view}
reveal that of the five reconstructed plumes only plume 6 lies in front of
the solar limbs as seen by \emph{STEREO A} and \emph{B}, the footpoints of
the other four are hidden by the limb. By projecting the reconstructed
plumes onto two simultaneous EUVI images (see Figure \ref{fig:proj_euvi_0601})
and taking into account Figures \ref{fig:side_view} and \ref{fig:top_view},
the spatial relationship between plumes and bright points
can be inferred. The plume in front of both limbs, that is, plume 6 could be
associated with a very faint bright point. Plume 7 presents a nice example of
the importance of having 3D information when associating plumes with bright
points. If we consider only EUVI B, then plume 7 seems to be related to a
bright point right in front of the limb.
However, when we check both Figures \ref{fig:side_view} and
\ref{fig:top_view} we find this plume is rooted just behind the limbs seen
by EUVI A and B. Therefore, the association to the bright point near the limb
is probably spurious.
For plumes 8 and 9, it is difficult to reach a firm conclusion. We see
two bright points in both images close to the plume roots. The association is
possible if the bright point relevant to plume 8 is big enough and the height
of the bright point relevant to plume 9 is large enough that it could be seen
from EUVI A and B, even though it is behind the limb.

\subsection{Results combining stereoscopy and SUMER observations}
\label{sec:stereosc_SUMER}

For the data set in April, SUMER observations are available. To obtain the
electron density and electron temperature along a plume,
we assume that the geometry of a plume does not change during its evolution, and
then project it onto the density and temperature maps deduced from the line
ratio of the Si\,{\sc viii} line pair and the Mg\,{\sc ix} line pair,
respectively \citep{Wilhelm:2006, Wilhelm:etal:2009}.
Consistent with the previous results, the
plumes are denser and cooler than the interplume regions.
The Doppler shift measurements, we deduced from the O\,{\sc vi} lines
based on the method outlined in \citet{Wilhelm:etal:1998}.

\begin{figure} 
  \vbox{
  \includegraphics[width=8cm,height=4cm]{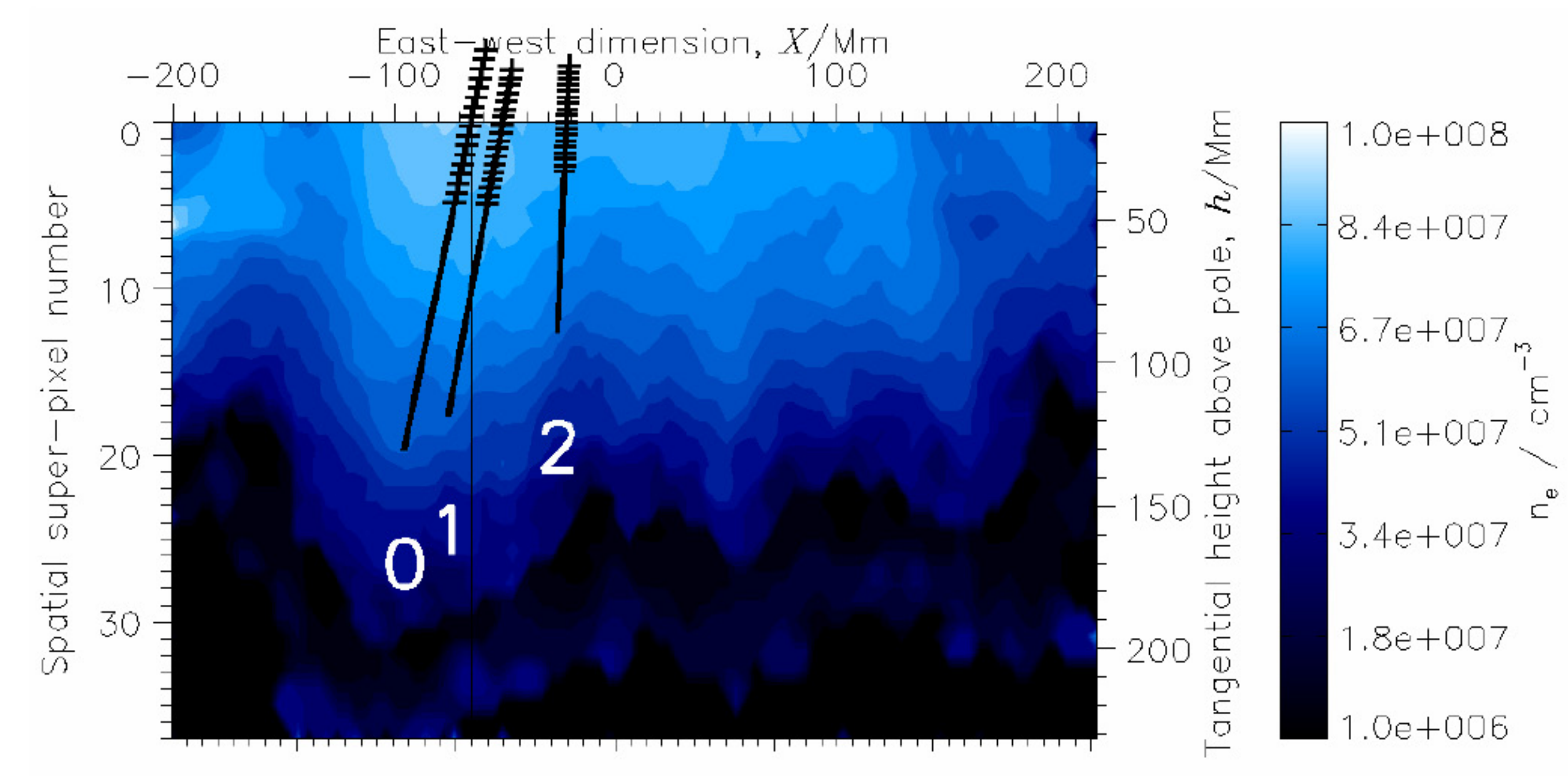}
  \includegraphics[width=8cm,height=4cm]{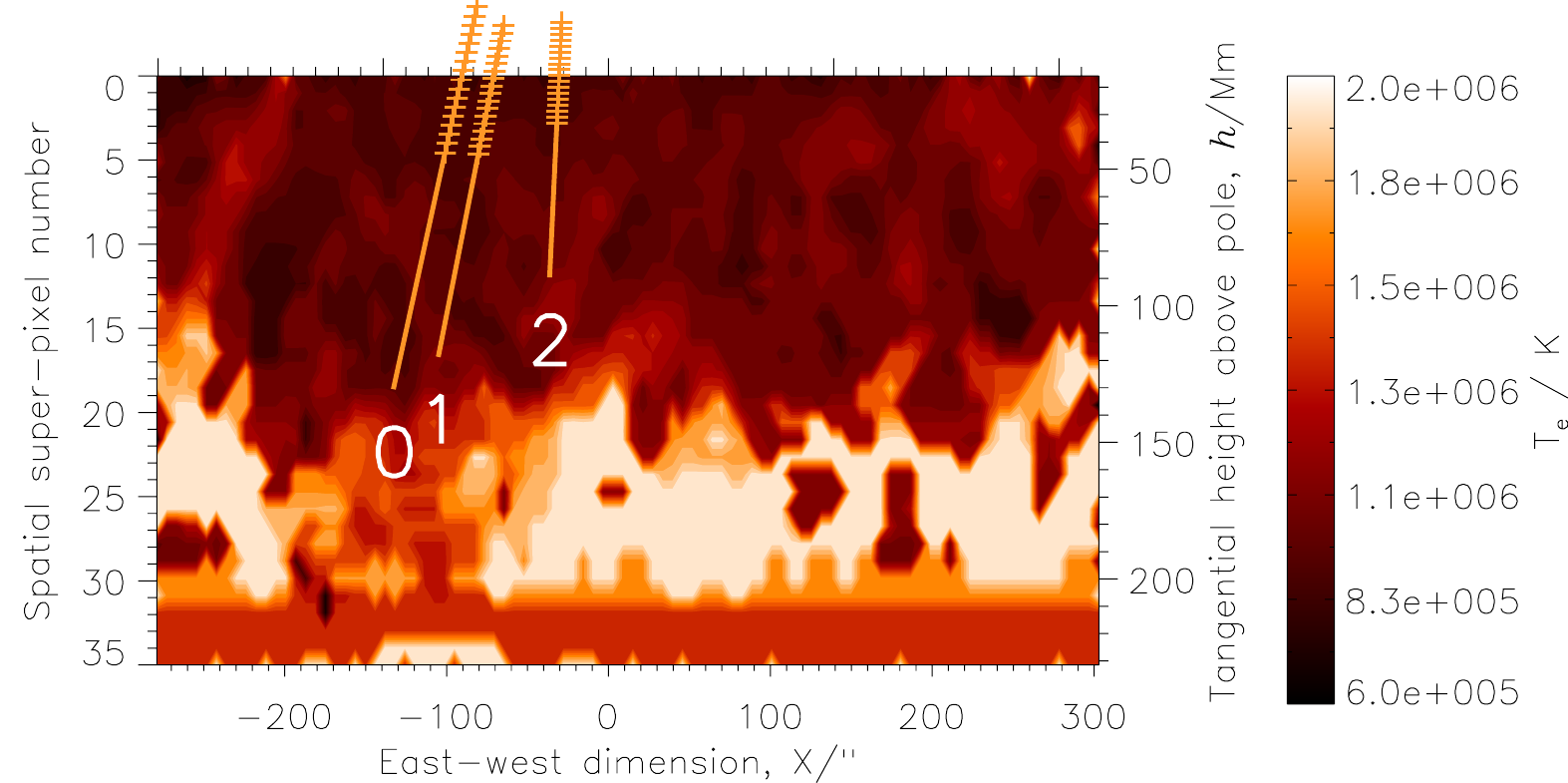}}
  \caption{Projections of the 3D plumes 0, 1 and 2 onto the SUMER electron
  density (upper) and
  temperature (lower) map with the marked plume numbers. The vertical solid
  line in the density map corresponds to the position of SUMER slit
  at April 8 01:00 UTC. The plus signs are the projections of the 3D
  plumes reconstructed from two EUVI images recorded on April 7 and solid lines are the
  extrapolations downwards to the point where the standard deviation of the
  temperature along the plume is around 0.1 MK. }
  \label{fig:sumer_dentem}
\end{figure}

SUMER scanned the relevant region from April 7 01:01 UTC to April 8 12:19 UTC
continuously moving from west to east. To compare these data with the EUVI
observations of a 3D plume, we need to first rotate the Sun from the EUVI
observations to the time at which SUMER scanned it. An example is shown in the density map
in the upper panel of Figure \ref{fig:sumer_dentem}. The inclined lines are
the projected positions of three plumes. The vertical line corresponds to the
position of the SUMER slit at April 8 01:00 UTC. Consider plume 1, the
plotted location was obtained by first rotating the Sun to this time and then
projecting plume 1 onto the density map. From Figure \ref{fig:sumer_dentem}
we can see that the projected plume 1 and the corresponding slit position are
consistent.

The time at which the SUMER slit passes through the centers of the three
plumes 0, 1 and 2 are April 8 02:00 UTC, April 8 01:00 UTC and April 7
22:00~UTC, is less than four hours after the EUVI observations. Due to the
inclination of these three plumes, they are scanned by SUMER for about 2.5~h,
2~h
and 1~h, respectively. We have checked the EUVI images at April 8 02:00 UTC,
April 8 01:00 UTC and found that plume 0 and 1 were still present though the
plumes appeared more diffuse at 02:00 UTC. For plume 3, we could not make a
comparison because
the time difference between the EUVI observation and the corresponding time
at which it was scanned by SUMER is too large.
Plume 4 in the EUVI observation lies outside the field of view of the SUMER
scan. In Figure \ref{fig:sumer_dentem}, the plus signs are the projections of the 3D plumes
reconstructed from two EUVI images, the solid lines are their extrapolations
outwards. We somewhat arbitrarily choose the upper end of height profile
where the temperature by SUMER does not dramatically deviate from 0.9~MK or so. It
makes the plumes more or less isothermal, which is an assumption for the later
calculations.
Furthermore, above $\approx 120$ Mm the temperature plot is too noisy for a
quantitative analysis.
For the density, $\approx 150$ Mm to 170 Mm might be an approximate limit.

We projected a 3D plume onto LOS Doppler shift map as well to get a more
precise outflow velocity along the plume by dividing the SUMER
velocity with the cosine of the plume inclination angle to the LOS.
However, we did not find any
significant Doppler shift from SUMER observations, the maximum of the Doppler
velocities is around 3 $\mathrm{km\,s}^{-1}$. If we take this number as a reference to estimate
the outflow velocities along the plumes 0, 1 and 2, we found that they are
quite small with a maximum of 10 $\mathrm{km\,s}^{-1}$. This
speed is much smaller than the sound speed $c_s \approx 140\,\mathrm{km~s^{-1}}$ for a
temperature of $\approx 0.9$ MK.

Similar to \citet{Gabriel:etal:2003}, we have made an estimate of
the plume contribution to the fast solar wind. The proton flux density for
the high-latitude fast solar wind observed during
the solar minimum from Ulysses at $r_{E}=1 \rm{AU}$ is 2.05 $\times 10^8$~
$\mathrm{cm}^{-2}\,\mathrm{s}^{-1}$ \citep{McComas:etal:2000}. We take the
cross sectional area of the coronal hole from \citet{Munro:Jackson:1977} that
matches the observed values extremely well,
\begin{equation}
A(r)=A_0(\frac{r}{r_0})^2f(r),
\end{equation}
where the subscript 0 refers to quantities evaluated near the solar surface
and $f(r)$ is the area expansion factor which reaches an almost constant value
of 7.26 beyond 3 $R_{\odot}$. Therefore, the mean proton flux density
mapped to the solar surface in the coronal hole is:
\begin{equation}
\frac{2.05\times10^8\times A(\mathrm{r_E})}{A_0}=2.05\times10^8
\times(\frac{\mathrm{r_E}}{r_0})^2\times7.26 \,\mathrm{cm}^{-2}\,\mathrm{s}^{-1}\approx 6.88\times10^{13}
\,\mathrm{cm}^{-2}\,\mathrm{s}^{-1}.
\end{equation}
On the other hand, the electron flux density contributed by plumes are estimated by
taking the maximal density in Figure 13 and maximal velocity of 10~$\mathrm{km\,s}^{-1}$, that
is, $10^{7.8}\,\mathrm{cm}^{-3} \times 10^6\,\mathrm{cm\,s}^{-1} \approx
6.31 \times 10^{13}\,\mathrm{cm}^{-2}\,\mathrm{s}^{-1}$.
Comparing the mean flux density and the flux density from plumes, we find
that the former is a little bit higher than the latter, and taking into
account the filling factor of plumes in coronal holes of 10~\%
\citep{Ahmad:Withbroe:1977}, it is unlikely
that the plumes investigated in this work are a dominant contributor to the
fast solar wind.

We assume that the plume plasma is in hydrostatic balance considering their
long lifetime of one or more days and the absence of any measurable flow.
However, if the plasma ions are heated by, e.g., ion cyclotron waves a
thermodynamic equilibrium does not necessarily exist since the plasma is
mainly cooled by a divergent electron heat flux and by inelastic electron
collisions with the ions \citep{Tu:Marsch:1997}.
Following their two-fluid approach, the sum $p=p_i+p_e$ of ion and electron
pressure has to obey
\begin{equation}
  \frac{d}{dr}(p_i+p_e)(r) = -m_i n g_\odot \frac{R^2_\odot}{r^2}
\end{equation}
Quasineutrality in this balance is insured by an ambipolar electric field
which cancels when the momentum equations for ions and electrons are added.
Here, $m_i$ is the mean ion mass, $n$ the plasma density and $g_\odot$ the
gravity acceleration at the solar surface.
For the total pressure we have
\begin{equation}
 p = p_i+p_e = n k_B (T_i+T_e) = 2n k_B T_\lambda
\end{equation}
Insertion and integration yields, changing the variable $r$ to $h=r-R_\odot$
\begin{equation}
  p = p_0 \exp {\left(-\frac{m_i g_\odot}{k_B}
                \int_0^h \frac{R^2_\odot}{T_\lambda
                (R_\odot+h')^2}\,dh'\right)}
\end{equation}
For the small height range of our observations, we can neglect a possible
height variation of $T_\lambda$ inside the plumes. Then
\begin{equation}
  n \simeq n_0 \exp {\left(-\frac{m_i g_\odot}{k_B T_\lambda}\;\frac{h}
         {(1+h/R_\odot)}\right)}
         = n_0 \exp {\left(-\frac{h}{\lambda_n(T_\lambda)\,(1+h/R_\odot)}\right)}
\label{equ:ne_h}
\end{equation}
where the scale height $\lambda_n(T_\lambda)$ depends on $T_\lambda=(T_i+T_e)/2$
and $n_{\rm 0}$ is the density at $h=0$ taken to be the base of the corona.
For typical coronal mean mass $m_i$ we have
$\lambda_n(T_\lambda) \simeq 47 \,\mathrm{Mm}\,(T_\lambda/\mathrm{MK})$
\citep{Aschwanden:2004}.

\begin{figure} 
  \vbox{
  \includegraphics[width=8cm,height=6.cm]{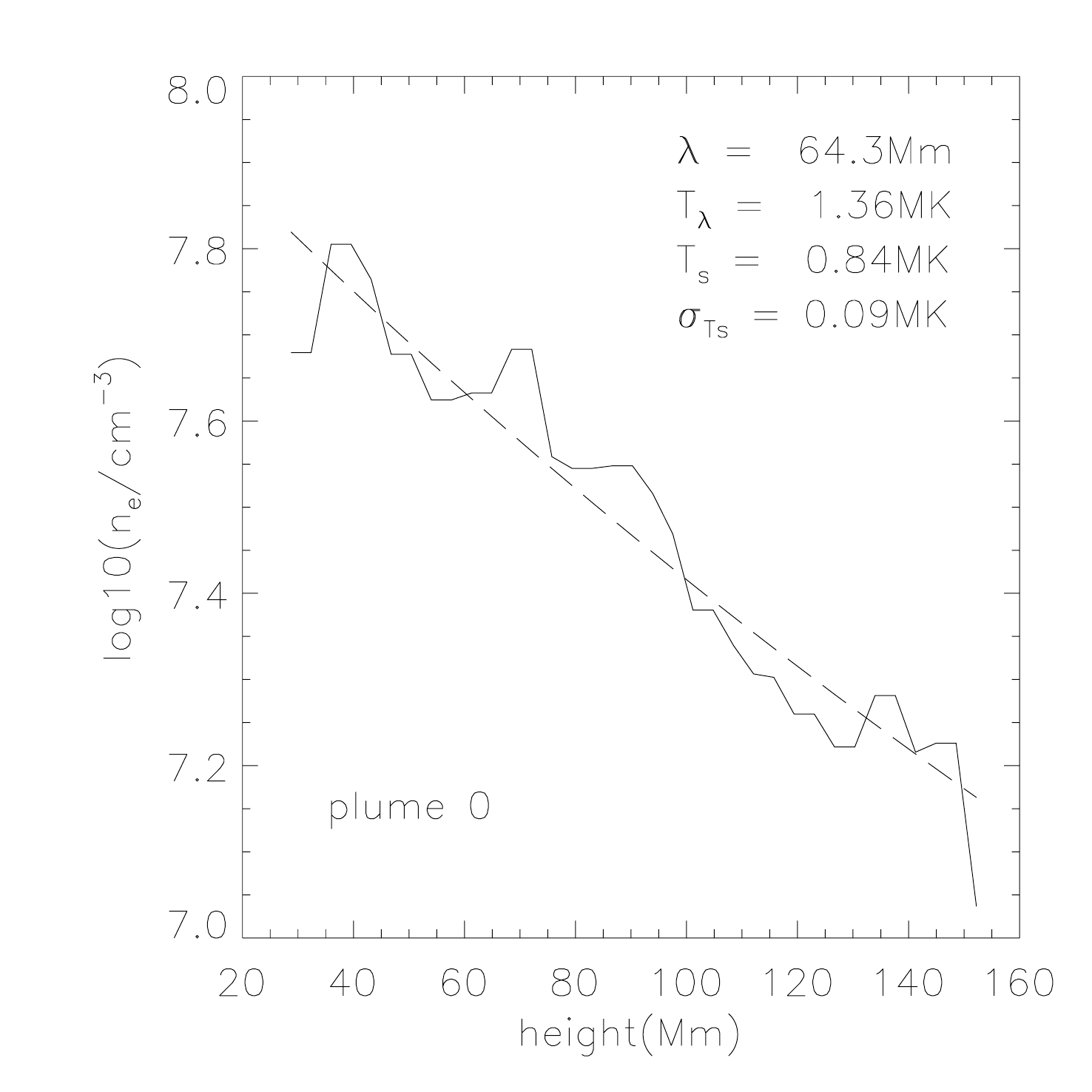}
  \includegraphics[width=8cm,height=6.cm]{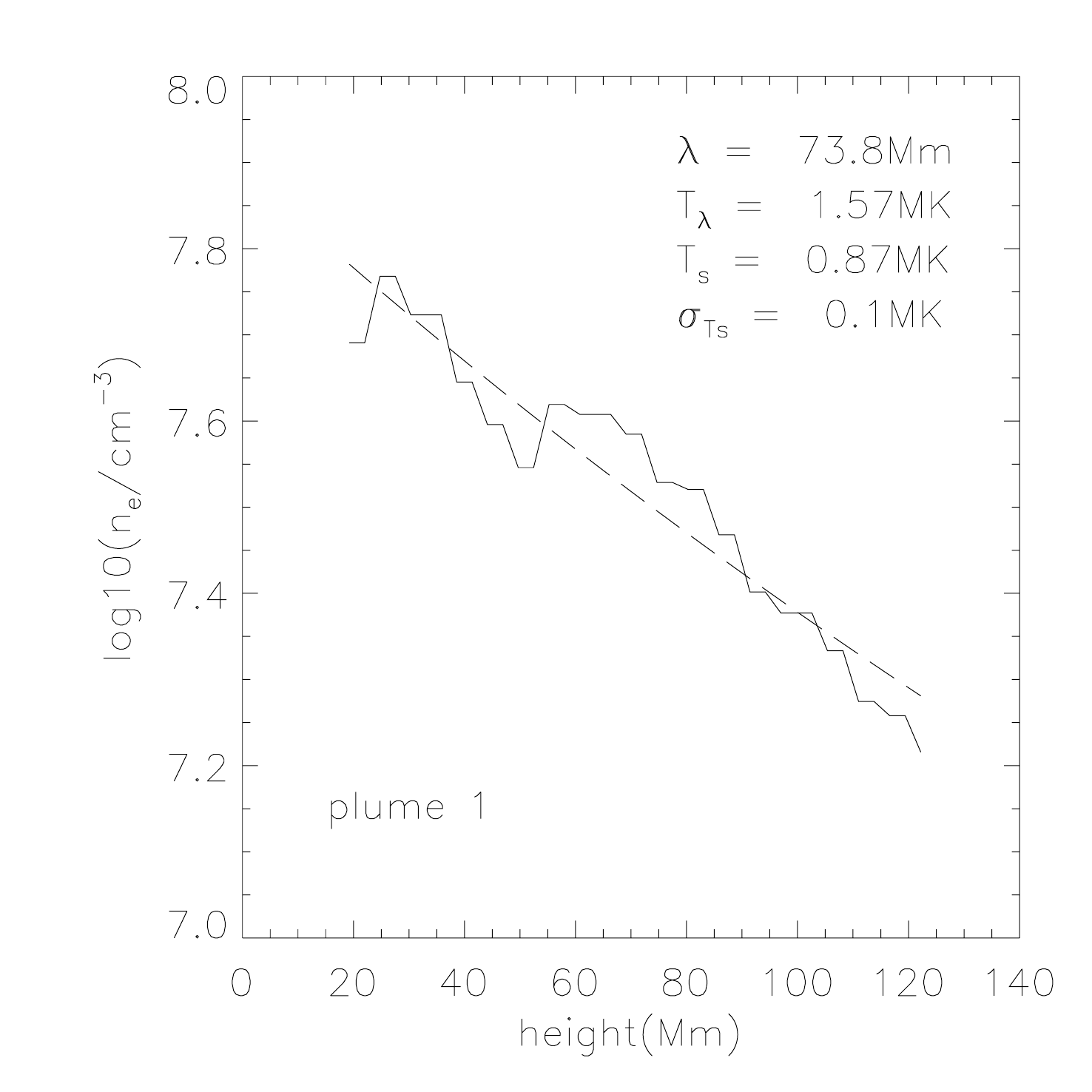}
  \includegraphics[width=8cm,height=6.cm]{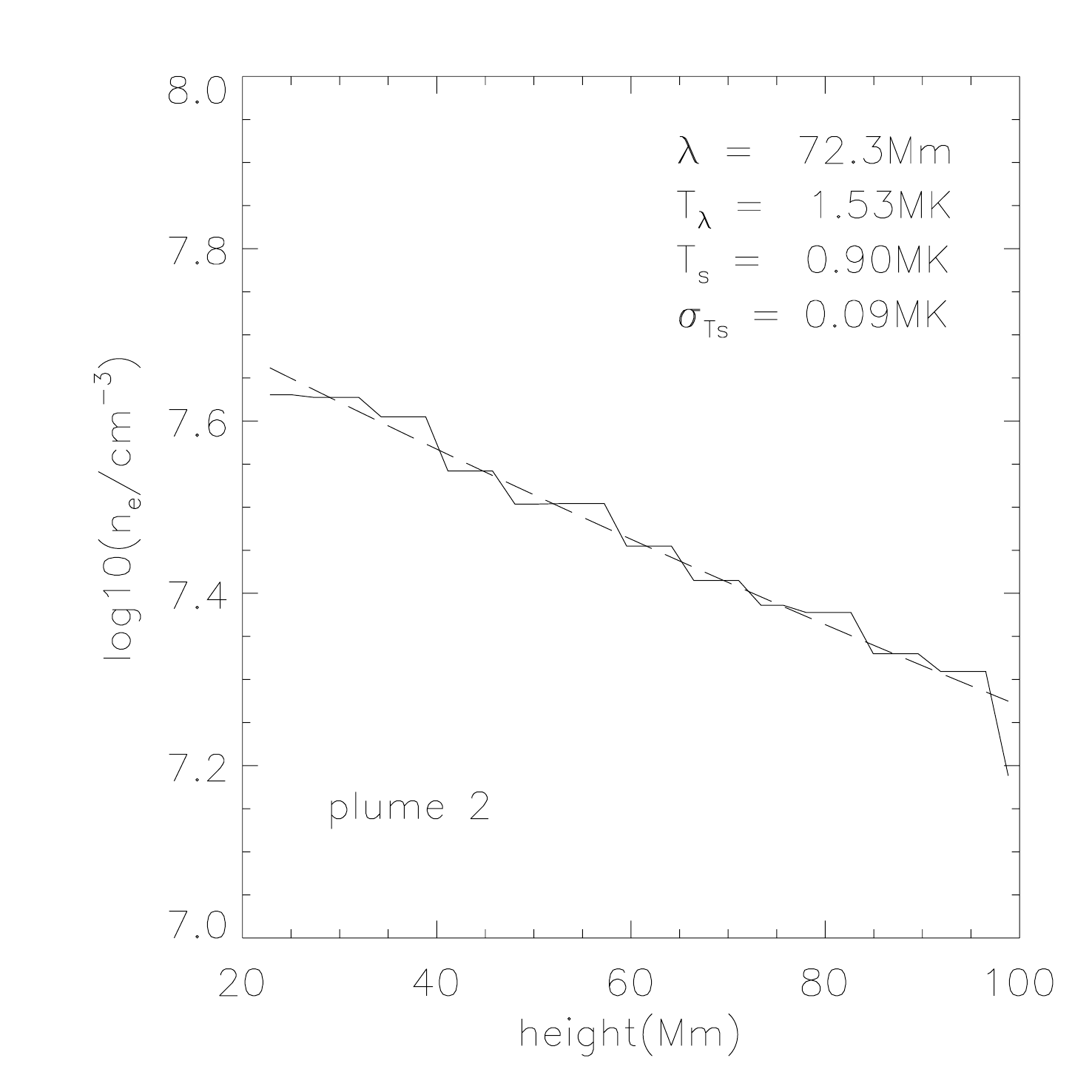}}
  \caption{Logarithmic electron density along the plumes 0, 1, 2 from
  SUMER observation as a function of height above the solar surface as deduced
  from stereoscopic reconstructions (Solid lines). The dashed lines are fits
  based on hydrostatic equilibrium described by Equation \ref{equ:ne_h}.
  In the upper right of each figure we mark the
  numerical values for $\lambda$ (density scale height), $T_\lambda$ (temperature
  corresponding to scale height), $T_s$ (electron temperature from SUMER) and
  its standard deviation $\sigma_{Ts}$.}
  \label{fig:sumer_fit}
\end{figure}

Since we know the density along the plumes from SUMER observations and we
know the 3D height from the stereoscopic reconstructions, we can fit these two
variables, $n_{\rm e}$ and $h$ to derive $T_{\lambda}$ and $n_0$ in
Equation \ref{equ:ne_h}. In Figure \ref{fig:sumer_fit}, we present the results
of fits based on Equation \ref{equ:ne_h} to the density stratification of
plumes 0, 1, 2. The calculated density scale height is given along with the
corresponding temperature $T_{\lambda}$ and the electron temperature deduced
from SUMER $T_{\rm s}$ in Figure~\ref{fig:sumer_fit}. We have invariably
$T_\lambda > T_{\rm s}$. The fits describe the data reasonably well suggesting that they
are consistent with hydrostatic equilibrium.

\section{Discussion and outlook}

We have reconstructed the three dimensional
geometry of ten polar plumes using simultaneous observations by the two
\emph{STEREO} spacecraft. For two different days, the locations of the footpoints of
ten
plumes and their inclinations were determined. Even though the statistical
basis is small, we find that the plumes we could detect from the EUV images
are homogeneously distributed over the polar cap. For both cases, the deviation of the plumes
to the local meridian plane is rather small with an average of $6.47^{\circ}$.
The deviation of the plume projection onto the local meridian plane
from the local radial direction becomes larger in general
with increasing distance of the plume from solar poles. For these two data sets,
a simple dipole model with its axis along the solar rotation axis for the
global magnetic field does, however, not provide a good description of the
obtained inclinations. The magnetic field in these two coronal holes were more
horizontal than this dipole field by $12.9^{\circ}$ on average. The lower the
latitude is, the larger is the deviation from the dipole field.

Moreover, we find that EUV plumes and EUV surface bright points
are not always related, which is consistent with the observations of
\citet{Wang:1998} and \citet{Raouafi:etal:2008}. Of the three plumes
in front of the solar
limb on April 7, only one was definitely associated with a
bright point. For the other two we did not find a related bright point.
Conversely, we saw a bright point in the images to which no
plume pair could be assigned to in the two EUVI
images. A possible explanation could be that the lifetime of a bright point
is shorter than the formation and decay time of a plume. \citet{Wang:1998}
assumes a bright-point lifetime of around 12 hours, while for plumes
he assumes lifetimes around one day.
From the case study of June, we find that care is required when paring
plumes with bright points. Spatial
coincidence in a single image could easily be misleading.

For the data set in April, based on the results of 3D reconstruction and
electron temperature, electron density and Doppler shift derived from the
SUMER observations, we calculated the density scale height
by assuming that a plume is in hydrostatic equilibrium. Using the reconstructed
3D direction of the plumes in space we could set an upper limit of 10
$\rm{km~s^{-1}}$ for the outflow speed along the plumes. The absence of a significant flow in plumes at
heights less than 1.2 $R_{\odot}$ is in agreement with the conclusions of
\citet{Wilhelm:etal:2000} and \citet{Raouafi:etal:2007}.
The temperatures derived from the density scale height were all in excess of
1~MK, while SUMER derived electron temperatures were well below 1~MK from line
ratios of Mg\,{\sc ix}. The ratio of the temperature obtained from the scale
height, $T_\lambda$, to the blue electron temperature from SUMER, $T_{\rm s}$, is
$T_\lambda/T_{\rm s}\approx$ 1.62 to 1.81. Recently,
\citet{DelZanna:etal:2008} found that the coronal electron temperatures derived from
the Mg\,{\sc ix} line ratio may have been significantly underestimated. A coronal hole inter-plume
temperature of 0.85 MK is now revised to 1.16 MK. This conclusion would
reduce the discrepancy between the temperatures derived from the two techniques
in our work, but would not eliminate it entirely. Even with this correction,
$T_\lambda/T_{\rm s}\approx$ 1.32 to 1.46 remains.
A possible explanation for this difference of $T_\lambda$ and $T_s$
could be a deviation of
electron and ion temperatures. The scale height depends on the average of
the electron and the ion temperature while the Mg\,{\sc ix} line-ratio depends
on the electron
temperature. The corrected ratio of $T_\lambda$ to
$T_{\rm s}$ corresponds to a ratio of the ion temperature to the electron
temperature of from 1.64 to 1.92, which is qualitatively consistent with
the result of \citet{Wilhelm:2006} derived from a different method. The
effective ion temperatures he deduced from the line widths are
higher than the electron temperatures as well.

This study has demonstrated that it is possible to uniquely obtain the
location and inclination of plumes from stereoscopic reconstructions. These
properties can significantly enrich our knowledge of plumes, in particular
when combined with other type of data, such as FUV spectra (as shown here) or
magnetograms (not shown here). Further stereoscopic reconstructions of a
large number of plumes would be very valuable.

\acknowledgements

The authors appreciated the constructive comments of the referee. \emph{STEREO}
is a project of NASA, \emph{SOHO} a
joint ESA/NASA project. LF and KW are members of an ISSI International Study
Team on plumes. LF was supported by the IMPRS graduate school run jointly
by the Max Planck Society and the Universities G\"ottingen and Braunschweig.
TW was supported by DLR grant 50OC0501 and WQG was supported
by National Basic Research Program of China (2006CB806302).

\begin{table*} 
\begin{tabular}[b]{l|cc|cc}
\hline
Date                       &\multicolumn{2}{|c|}{April 7, 2007} &\multicolumn{2}{c}{June 1, 2007} \\
\hline
STEREO spacecraft           &B           &A                &B             &A\\
\hline
\hline
Sun-Earth. dist.(d/au)       &1.0277777     &0.9640101      &  1.0647322   &0.95860281\\
Sun's app. rad.($R_{\odot}/"$)    &933.69200     &995.45200      &901.28300     &1000.3757\\
HEEQ longitude($\varphi/^\circ$)     &-0.918945     &2.697643       &-3.6847666    &6.7788499\\
HEEQ latitude($\theta/^\circ$)      &-6.404608     &-5.999570      &-1.4801556    &0.2276848 \\
resolution($\Delta x/(^{\prime\prime}/pix)$)&1.5900000     &1.5877700      &1.5900000     &1.5877740   \\
Separation ($\alpha/^\circ$)        &\multicolumn{2}{|c|}{3.62}    & \multicolumn{2}{c}{10.6}   \\
\hline
\hline
Exposure time (T/s)     &\multicolumn{2}{|c|}{20}      &\multicolumn{2}{c}{16}    \\
Compression mode           &\multicolumn{2}{|c|}{ICER4}  & \multicolumn{2}{c}{ICER4}   \\
\hline
\end{tabular}
\caption{The positions of the two \emph{STEREO} spacecraft given in HEEQ coordinate
system, exposure time of the observation and compression mode of the recorded
data on 2007-04-07 22:01:17 UTC and on 2007-06-01 00:09:00 UTC. }
\label{tab:orbit_plume}
\end{table*}

\begin{table*} 
\begin{tabular}{c|ccccc|ccccc}
\hline
Date                   &\multicolumn{5}{|c|}{April 7}
                       &\multicolumn{5}{c}{June 1}\\
\hline
Plume                  &  0  &   1  &    2 &   3  &  4
                       & 5    & 6   & 7    & 8   &9 \\
\hline
$\lambda/^\circ$       &-84.7& -85.0& -78.0& -85.4& -69.7
                       &74.5  &83.2 &86.9  &84.4 &79.6\\
$\varphi/^\circ$       &-69.9&-131.3&-171.6& 96.9 & 132.5
                       &-145.6&-60.7&-177.9&111.7&141.3\\
$\beta/^\circ$         & 65.7& 106.7& 117.4&  94.4& 128.6
                       &118.8 &68.6 &103.2 &97.6 &114.8\\
$\gamma/^\circ$        & 12.2& -8.61 & -1.70 & 8.67 &9.85
                       &-2.25 & 14.2& -4.48&1.10 &1.62\\
$\psi/^\circ$          &-15.0  &-19.9 & -21.9 &-10.4 &-37.0
                       &17.4  &13.2 &9.35  &10.4 &19.6 \\
$\tan(i)$              &-3.73 & -2.77 & -2.52 & -5.45 & -1.32
                       &3.19  &4.26 &6.07  &5.43 &2.80 \\
$2\tan(\lambda)$  &-21.9 &-22.3 &-9.40 & -24.9& -5.42
                       &7.20  &16.7 &37.5 &20.4 &10.9\\
$(i-i^{\prime})/^\circ$&12.4  &17.3 &15.8 &8.10 &26.5
                       &-9.49 &-9.77&-7.82&-7.59&-14.4\\
\hline
\end{tabular}
\hspace*{0.1pt}
\caption{Footpoint position and inclination of the reconstructed plumes
on April 7 and on June 1. The relevant angle symbols are the same as defined
in Figure~\ref{fig:plume_geometry}. The first two rows are the footpoint
latitudes and longitudes of the plumes. $\beta$ gives the angle between a plume
orientation and the LOS direction of the Earth. $\gamma$ measures how much a
plume deviates from the local meridian plane and its sign depends on the
sign of ${\widehat{\mathbf{e}}_{plume}}~^T\cdot \widehat{\mathbf{e_\varphi}}$.
$\psi$ is the complementary angle of magnetic inclination $i$ and its sign is
the same as the sign of
${\widehat{\mathbf{e}}_{plume}}~^T\cdot \widehat{\mathbf{e_\theta}}$.
All angles are given in units of degrees.
The last three rows are the tangent of magnetic inclination, two times
the tangent of the latitude, the difference of the magnetic inclination $i$
with the dipole magnetic inclination $i^{\prime}$ corresponding to
$\arctan(2\tan\lambda)$.}
\label{tab:parameter}
\end{table*}

\end{document}